\documentclass[final,5p,times,twocolumn
]{elsarticle}

\usepackage{alltt}  
\usepackage{amsmath}
\usepackage{graphicx}
\usepackage{subcaption}
\usepackage{url}
\usepackage{color}
\usepackage{float}
\usepackage{etoolbox,siunitx}
\usepackage{booktabs}
\usepackage[acronyms]{glossaries}
\usepackage{mhchem}
\usepackage{amssymb}
\usepackage{my_paper}
\usepackage{colortbl}
\usepackage[pdftex,colorlinks=true, linkcolor = blue, citecolor=blue,urlcolor=blue, bookmarksnumbered=true, bookmarksopen=true]{hyperref}
\usepackage{cleveref}
\usepackage{widetext}
\usepackage{flushend}
\usepackage{cuted}

\usepackage{tikz}

\usetikzlibrary{shapes.geometric, arrows}

\tikzstyle{startstop} = [rectangle, rounded corners, minimum width=3cm, minimum height=1cm,text centered, draw=black, fill=red!30]
\tikzstyle{process} = [rectangle, minimum width=3cm, minimum height=1cm, text centered, draw=black, fill=orange!30]
\tikzstyle{decision} = [diamond, minimum width=3cm, minimum height=1cm, text centered, draw=black, fill=green!30]
\tikzstyle{arrow} = [thick,->,>=stealth]
\tikzstyle{io} = [trapezium, trapezium left angle=70, trapezium right angle=110, minimum width=3cm, minimum height=1cm, text centered, draw=black, fill=blue!30]

\crefname{table}{Table}{Tables}
\crefname{figure}{Fig.}{Figs.}
\crefname{equation}{Eq.}{Eqs.}

\newcounter{bla}

\journal{Computer Physics Communications}
\hypersetup{draft}
\begin{document}

\begin{frontmatter}



\title{The LISE package: solvers for static and time-dependent superfluid local density approximation equations in three dimensions}


\author[a]{Shi Jin}
\author[b,a]{Kenneth J. Roche}
\author[c]{Ionel Stetcu\corref{author}}
\author[a]{Ibrahim Abdurrahman}
\author[a]{Aurel Bulgac}

\cortext[author] {Corresponding author.\\\textit{E-mail address:} stetcu@lanl.gov}
\address[a]{Department of Physics,
  University of Washington, Seattle, Washington 98195--1560, USA}
\address[b]{Pacific Northwest National Laboratory, 
  Richland, Washington 99352, USA}
\address[c]{Theoretical Division, Los Alamos National Laboratory, 
  Los Alamos, New Mexico 87545, USA}

\begin{abstract}
%
%
%

Nuclear implementation of the density functional theory (DFT) is at present the only microscopic 
framework applicable to the whole nuclear landscape. The extension of DFT 
to superfluid systems in the spirit of the Kohn-Sham approach, 
the superfluid local density approximation (SLDA) and its 
extension to time-dependent situations, time-dependent superfluid local density 
approximation (TDSLDA), have been extensively used to describe various static and dynamical 
problems in nuclear physics, neutron star crust,  and cold atom systems. In this paper, we present the codes 
that solve the static and time-dependent SLDA equations in three-dimensional 
coordinate space without any symmetry restriction. These codes are fully 
parallelized with the message passing interface (MPI) library and take advantage 
of graphic processing units (GPU) for accelerating execution. The dynamic codes have 
checkpoint/restart capabilities and for initial conditions one can use any generalized 
Slater determinant type of wave function.
The code can describe a large number of physical problems: nuclear fission, collisions 
of heavy ions, the interaction of quantized vortices with nuclei in the nuclear star crust,
excitation of superfluid fermion systems by time dependent external fields,
quantum shock waves, domain wall generation and propagation, the dynamics of 
the Anderson-Bogoliubov-Higgs mode, dynamics of fragmented condensates, vortex rings dynamics,
generation and dynamics of quantized vortices, their crossing and recombinations and 
the incipient phases of quantum turbulence.

\end{abstract}

\end{frontmatter}



{\bf PROGRAM SUMMARY/NEW VERSION PROGRAM SUMMARY}

\begin{small}
\noindent
{\em Program Title: LISE }                                          \\
{\em Licensing provisions: Standard CPC license }                                   \\
{\em Programming language:} C, CUDA                          \\

{\em Nature of problem:}\\
 The full description of nuclear fission and nuclear reactions within the mean field approximation 
 in real time within 
 the extension of the density functional theory to superfluid systems is an extremely 
 computationally demanding problem, which requires the solutions of a very large 
 system of nonlinear coupled complex partial differential equations in 3+1 coordinates. 
 Similar problems also appear in the case of cold atoms and in the dynamics of the 
 neutron star crust, which have been tackled within the same framework with the same codes. \\
 
{\em Solution method:}\\
 The evolution equations are discretized on a 3-dimensional spatial lattice and propagated in time. 
 Spatial derivatives are evaluated using the fast Fourier transform technique. The propagation in time is performed
 using a predictor-modifier-corrector algorithm due to Adams-Bashforth-Milne, which requires 
 only  two evaluations of the right hand side of the equations per time step. The 
 accuracy of the time integration is $\sim{\cal O}(\Delta t)^6$. This method has a low truncation error, 
 excellent numerical stability,  and low roundoff errors.\\
 
{\em Additional comments:}\\
 The code has been implemented on a variety of supercomputers (Jaguar, Titan, Piz Daint, Tsubame, Summit, Sierra) 
 and demonstrates excellent scaling properties. In strong scaling to a large number of GPUs, the communication time
 between processes overtakes the computation time as the dominant run time cost. 
 \\

\end{small}



\section{Introduction} \label{sec:intro}

Density functional theory (DFT) and other self-consistent approaches
like Hartree-Fock (HF), Hartree-Fock-Bogoliubov (HFB), or Hartree-de
Gennes have played an essential role in studying the properties of
most nuclei across the nuclear chart, as well as for neutron star
crust and cold atom properties. We will restrict here the description 
of red our time-dependent code only for the case of nuclear fission and nuclear collisions.
The use of the code to describe neutron star crust or cold atom dynamics
is rather straightforward. The code and further version can be downloaded from on GitHub~\cite{github}. 

Present phenomenological nuclear
energy density functionals (NEDF) allow for a quite accurate description of
many bulk properties of nuclei such as masses, radii and shapes,
transition matrix elements, potential energy surfaces and related
inertial parameters, and even non-equilibrium properties. The
time-dependent extension of DFT is straightforward and widely used in
studying various nuclear dynamics, e.g. giant resonances, collisions
and fission~\cite{Maruhn:2014,Umar:2017,Simenel:2016,Tanimura:2015,
tanimura2017,Goddard:2015,Goddard:2016,
Scamps:2015,Bulgac:2016,Stetcu:2015,Stetcu:2011,Bulgac:2019b,Bulgac:2020}.

The time-dependent superfluid local density approximation (TDSLDA), 
which formally appears like the time-dependent Hartree-Fock-Bogoliubov  (TDHFB) 
approximation (in practice typically without any non-local Fock terms however) 
overcomes limitation of TDHF,
by treating explicitly the dynamics of the pair correlations. The acronym SLDA is a
natural extension of the Kohn-Sham acronym for the local density 
approximation (LDA) to superfluid systems.
In typical nuclear implementations of HF and HFB the non-local Fock terms are
localized, using various approximations, and the Fock designation is typically superfluous. 
The only exception is when one uses the Gogny interaction, which is treated formally 
as a ``real'' nucleon-nucleon interaction and the emerging equations are non-local. 
For electronic systems with their long range Coulomb interaction there was never a need to resort 
to non-local meanfield equations within DFT and no theoretical argument was ever 
made that an explicit treatment of non-locality is indeed necessary in nuclear physics, 
where the range of the nucleon-nucleon interaction is comparable to the average 
nucleon-nucleon separation.  The only remnant of the finite-range  character of the nuclear forces
is manifest in the appearance of an effective nucleon mass~\cite{Negele:1972}.
In modern nuclear meanfield implementations 
one typically starts from a local energy density functional
and the emerging equations look either like Hartree or Hartree-Bogoliubov-de Gennes 
equations, with a coordinate dependent effective mass. 
Since the NEDF is not obtained as an expectation of the nuclear Hamiltonian
over a generalized Slater determinant, we prefer to use the terms LDA and their natural 
generalizations in the spirit of various Kohn-Sham incarnations of the DFT~\cite{Nakatsukasa:2016,Schunck:2019,Colo:2020}.
The (TD)SLDA is formulated in terms of quasiparticle wave functions, discretized on
 3D spatial lattice of size $(N_x, N_y, N_z)$. The number of quasiparticle wavefunctions
(qpwfs) is comparable to the dimension of the quasiparticle hamiltonian $N =
4N_xN_yN_z$, which could reach ${\cal O}(10^6)$ in simulations for heavy
nuclei, see \cref{sec:theo}.
In TDHF simulations of nuclear systems, the number of sp wavefunctions is
comparable to the number of nucleons, which is typically of the order of a few hundred orbitals.

The initial conditions of TDHF(B) equations are prepared typically by solving the
self-consistent static HF(B) equations with appropriate constraints. In static HF(+BCS)
calculations, the solution is usually obtained by performing imaginary
time evolution thanks to the smaller number of single particle
wavefunctions, such as in the EV8~\cite{Ryssens:2015} and Sky3D codes~\cite{Maruhn:2014}. The typical
implementation of static HF(+BCS) solvers involves iterative direct
diagonalizations of HF(+BCS) hamiltonian matrix, which can be divided into
two main classes. In the first one, the HFB problem is formulated in
the configuration space by expanding the quasiparticle states of HFB
on a discrete basis of orthogonal functions, usually provided by a
(deformed) harmonic oscillator (HO) basis~\cite{dobaczewski1997,Perez:2017,Schunck:2017}. 
Although typically very fast and accurate at deformations smaller 
than those corresponding to scission configurations, this approach 
suffers from truncation errors that typically lead
to  inaccurate description of the asymptotic behavior of the system
and practically it is impossible to describe separated fission fragments 
at scission and beyond.
Another approach is the direct HFB matrix diagonalization in
the coordinate space. Quite a number of
coordinate-space HFB solvers have been published over the years, but
solving the HFB equations in full 3D coordinate space is still very
computational expensive due to the large dimension $N$ of the HFB
matrix discretized in a large box. In an earlier work \cite{jin2017}, 
recently  extended to finite temperatures~\cite{Kashiwaba:2020},
we proposed an alternative approach to solve the SLDA/HFB equations based on a
Krylov subspace method, which eschews the need for diagonalization, 
but this approach cannot generate the qpwfs needed in TDSLDA/TDHFB
simulations.

The initial conditions for the TDSLDA problem are typically prepared from 
static SLDA solutions. In case of fission the initial nuclear configuration 
is obtained from an SLDA  plus appropriate proton and neutron numbers and 
quadrupole and octupole constraints. In the case of the collision of two nuclei, the initial target 
and the projectile are separately prepared in their ground states and 
subsequently  arranged at a suitable separation in larger simulation box.
The qpwfs of the entire projectile+target system are obtained by 
diagonalized the combined quasiparticle Hamiltonian. Since the Coulomb 
interaction is long ranged, it affects the chemical potentials of the protons in
both the target and projectile nuclei. Once these qpwfs of the combined 
system are determined both nuclei are given a boost. We make sure that 
during the simulations the center of mass coordinate is fixed and that 
the long axis of the simulation box is always aligned
with the instantaneous axis of the largest quadrupole moment of 
the combined system.

In this paper we present the codes needed to perform both static SLDA and dynamic 
TDSLDA calculations in an unified framework. The static code is a pure
CPU code. We solve the self-consistent SLDA equations by diagonalizing
the quasiparticle Hamiltonian with the ScaLAPACK library. The quasiparticle
Hamiltonian is distributed over different CPU processes. The
generated qpwfs are written onto disk, 
to be subsequently used by the TD code, which uses GPUs to further accelerate large
parts of the computation. The large-scale system of coupled  
partial differential equations (PDEs) are solved with the
predictor-modifier-corrector Adams-Bashforth-Milner (ABM) method
\cite{Hamming:1996} with a total error $\propto (\Delta t^6)$ per time
step, associated with a truncated, but accurate series expansion of
evolution operator method for start/restart procedure.  All spatial
derivatives are evaluated using fast Fourier transforms (FFT). 
 These codes have been compiled and run on
many leadership supercomputers around the world, e.g. OLCF Titan and 
Summit (Oak Ridge); Piz Daint (Lugano, Switzerland), 
Tsubame (Tokyo Institute of Technology), Lassen and Sierra (Lawrence Livermore National Laboratory), 
Kodiak/Moonlight (Los Alamos National Laboratory).

\section{Theoretical Framework} \label{sec:theo}

\subsection{Nuclear DFT and Superfluid Local Density Approximation (SLDA)}

The NEDF in case of SLDA is formulated in terms of various densities, constructed from quasiparticle wave functions
\begin{equation}
\psi_k(\vec r)=\left(\begin{array}{c}
\text{u}_{k\uparrow}(\vec r)\\
\text{u}_{k\downarrow}(\vec r)\\
\text{v}_{k\uparrow}(\vec r)\\
\text{v}_{k\downarrow}(\vec r)
\end{array}\right). 
\end{equation}
The ground-state energy is calculated from a density functional $\mathcal{E}$, which depends on various 
normal and anomalous number densities~\cite{Bender:2003,Bulgac:2013,Bulgac:2019} by imposing a minimization condition
(with implied introduction of appropriate Lagrange multipliers to enforce the orthonormality of the qpwfs):
\begin{equation}
\frac{\delta \mathcal{E}}{\delta \phi^*_{k}(\vec r)}=0,
\label{eq:minDFTeq}
\end{equation}
where $\mathcal{E}$ is a functional which
depends on densities and currents. In particular,  $\mathcal{E}$ is a function of  the number density $\n(\vec{r})$, the kinetic density $\tau(\vec{r})$, the anomalous density $\kappa (\vec{r})$, the spin density $\vec{s}(\vec{r})$, the spin-current density $\vec{J}(\vec{r})$, and the current density $\vec{j}(\vec{r})$, which are obtained from different components of the quasiparticle wavefunctions $\phi_k(\vec r)$ as follows:
\begin{subequations}\label{eq:local_densities}
\begin{align}
\n(\vec{r}) &= \sum_{k,\sigma}  \text{v}^*_{k,\sigma}(\vec{r}) \text{v}_{k,\sigma}(\vec{r}) \label{eq:rho_t},\\
\kappa(\vec{r}) &= \sum_k \text{v}^*_{k\downarrow}(\vec{r})\text{u}_{k\uparrow}(\vec{r}),  \label{eq:nu_t}\\
\tau(\vec{r}) & = \sum_{k,\sigma}  \vec{\nabla}\text{v}^*_{k,\sigma}(\vec{r}) \cdot \vec{\nabla}\text{v}_{k,\sigma}(\vec{r}),   \label{eq:tau_t}\\
\vec{s}(\vec{r}) &= \sum_{k,\sigma,\sigma'} \vec{\sigma}_{ss'} \text{v}^*_{k,\sigma}(\vec{r})\text{v}_{k,s'}(\vec{r}), \label{eq:sod_t} \\
\vec{J} (\vec{r}) & = \left. \frac{1}{2i} (\vec{\nabla} - \vec{\nabla}^\prime) \times \vec{s}(\vec{r}, \vec{r}^\prime) \right |_{\vec{r}=\vec{r}^\prime},   \label{eq:J_t}\\
\vec{j}(\vec{r}) &= \frac{1}{2i} \sum_{k,\sigma} \left[ \text{v}_{k,\sigma}(\vec{r}) \vec{\nabla} \text{v}^*_{k,\sigma}(\vec{r}) 
                                                                      - \text{v}^*_{k,\sigma}(\vec{r}) \vec{\nabla} \text{v}_{k,\sigma}(\vec{r}) \right], \label{eq:j}
\end{align}
\end{subequations}
where $\sigma=\uparrow,\downarrow$ and the sums run over eigenstates of \cref{eq:hfb} with positives eigenvalues only $E_k>0$.

A generic nuclear energy density functional (NEDF) is represented as a sum of the kinetic 
$\mathcal{E}_{\mathrm{kin}}$, the interaction
$\mathcal{E}_{\mathrm{interaction}}$, the Coulomb
$\mathcal{E}_{\mathrm{Coul}}$, and the pairing
$\mathcal{E}_{\mathrm{pair}}$ contributions
\begin{align} \label{eq:totalenergy}
\mathcal{E} = \mathcal{E}_{\mathrm{kin}} + \mathcal{E}_{\mathrm{interaction}} + \mathcal{E}_{\mathrm{Coul}} + \mathcal{E}_{\mathrm{pair}}.
\end{align}
The kinetic component is simply given in terms of proton and neutron kinetic densities
\begin{align}\label{eq:kinenergy}
\mathcal{E}_{\mathrm{kin}}(\vec{r}) = \sum_{q=n,p} \frac{\hbar^2}{2m_q(\vec{r})}\tau_q (\vec{r}),
\end{align}
where in case of the Skyrme family of NEDFs one introduces a coordinate dependent nucleon effective mass $m_q(\vec{r})$.

\noindent
For protons, the Coulomb contribution to the density energy functional is composed of a direct and an exchange tem, the later being calculated in the Slater approximation:

\begin{align} \label{eq:coulomb}
\begin{split}
&\mathcal{E}_{\mathrm{Coul}}(\vec{r})  = \mathcal{E}^d_{\mathrm{Coul}}(\vec{r}) + \mathcal{E}^e_{\mathrm{Coul}}(\vec{r}) \\
& = \frac{e^2}{2} \int \frac{\n_p(\vec{r}) \n_p(\vec{r}_1)}{\lvert \vec{r} - \vec{r}_1 \rvert} d^3\! {r}_1- \frac{3e^2}{4} \left(\frac{3}{\pi} \right)^{1/3} \n_p^{4/3}(\vec{r}).
\end{split}
\end{align}
The pairing energy  depends on the local anomalous density:
\begin{align}\label{eq:pairing}
\mathcal{E}_{\mathrm{pair}}(\vec{r}) = \sum_{q=n,p} g_{\mathrm{eff}}(\vec{r}) \lvert \kappa_q(\vec{r}) \rvert^2
\end{align}
and the effective pairing coupling strength
$g_{\mathrm{eff}}(\vec{r})$ is obtained via a renormalization
\cite{bulgac2002, Yu:2003, Borycki:2006} of the bare pairing strength,
typically parametrized as
\begin{align}\label{eq:g0}
g_0 (\vec{r}) = g_0 \left ( 1 - \alpha \frac{\n(\vec{r})}{\n_0} \right ),
\end{align}
where $\n_0 = 0.16~\mathrm{fm}^{-3}$ is the saturation
density. The parameter $\alpha = 0, 1, 1/2$ corresponds to volume, surface, and
mixed pairing respectively \cite{Dobacz:2002, Bertsch:2009}.

The interaction part is complicated, as it needs to describe all correlations
induced by the underlying nucleon-nucleon interaction \cite{Hohenberg:1964}. Over the years,
different forms of the nuclear functional have been proposed; see Refs.~ \cite{Bender:2003,Dutra:2012} 
for reviews. Numerical solutions are significantly less demanding if the NEDF is local, hence 
the Skyrme family of NEDFs~\cite{Bender:2003,Dutra:2012} are rather popular. 
Such energy functionals have a generic form
\begin{align}\label{eq:skyrme}
\begin{split}
\mathcal{E}_{\mathrm{Skyrme}} & = \mathcal{E}_{\n^2} + \mathcal{E}_{\n^\gamma} + \mathcal{E}_{\n \Delta \n} + \mathcal{E}_{\n \tau}  + \mathcal{E}_{\n \nabla J} \\
& = \sum_{t=0,1}  C_t^{\n} \n_t^2   + C_t^\gamma \n_t^{2}\n_0^\gamma  + C_t^{\n \Delta \n} \n_t \Delta \n_t \\
 & + C_t^\tau (\n_t \tau_t - \vec{j}_t\cdot \vec{j}_t)  + C_t^{\nabla J}\left(\n_t \vec{\nabla} \cdot \vec{J}_t + \vec{s}_t\cdot (\vec{\nabla} \times \vec{j}_t)\right) 
\end{split}
\end{align}
where $\n_0=\n_n+\n_p$ and $\n_1=\n_n-\n_p$ (and similar
for $\tau_{0,1}$ and $\vec{J}_{0,1}$) are isoscalar and isovector number
densities respectively, and $C$'s are coupling constants.
In recent years, we have also developed a qualitatively new NEDF named SeaLL1 \cite{Bulgac:2018a}, which has a similar form to Skyrme NEDFs 
\begin{align}\label{eq:seaLL1}
\begin{split}
\mathcal{E}_{\mathrm{SeaLL1}} & = \mathcal{E}_{\mathrm{vol}} + \mathcal{E}_{\n \Delta \n}  + \mathcal{E}_{\n \nabla J} \\
& = \sum_{j=0}^2 (a_jn_0^{5/3} + b_j n_0^{2} + c_j n_0^{7/3}) \left( \frac{n_1}{n_0} \right)^j  \\
 & +  \sum_{t=0,1} C_t^{\n \Delta \n} \n_t \Delta \n_t + C_t^{\nabla J} \left(\n_t \vec{\nabla} \cdot \vec{J}_t + \vec{s}_t\cdot (\vec{\nabla} \times \vec{j}_t)\right) 
\end{split}
\end{align} 
with the coefficients $a_j, b_j, c_j$ and $C$'s specified in Ref. \cite{Bulgac:2018a}. SeaLL1 depends only on seven parameters, has an effective nucleon mass equal to the bare nucleon mass, and even it is not optimized yet, has a superior accuracy to \textit{any} Skyrme NEDFs. 

The minimization condition (\ref{eq:minDFTeq}) translates into  self-consistent eigenvalue equations, which by design are similar in form, 
but not  in their physical interpretation~\cite{Schunck:2019,Colo:2020,Bulgac:2013,Bulgac:2019},
with the local Hartree-Fock-Bogoliubov or Hartree  de Genes equations~\cite{Ring:2004} 
for the $\text{u}_{k,\sigma}(\vec{r})$ and $\text{v}_{k,\sigma}(\vec{r})$  components of qpwfs 
\begin{widetext}
\begin{gather}\label{eq:hfb}
\begin{pmatrix}
h_{\uparrow \uparrow}(\vec{r})-\mu  & h_{\uparrow \downarrow}(\vec{r}) & 0 & \Delta(\vec{r}) \\
h_{\downarrow \uparrow} (\vec{r})& h_{\downarrow \downarrow}(\vec{r}) - \mu& -\Delta(\vec{r}) & 0 \\
0 & -\Delta^*(\vec{r}) &  -h^*_{\uparrow \uparrow}(\vec{r}) +\mu & -h^*_{\uparrow \downarrow}(\vec{r}) \\
\Delta^*(\vec{r}) & 0 & -h^*_{\downarrow \uparrow}(\vec{r}) & -h^*_{\downarrow \downarrow}(\vec{r}) + \mu
\end{pmatrix}
\begin{pmatrix}
\text{u}_{k\uparrow}(\vec{r}) \\
\text{u}_{k\downarrow}(\vec{r}) \\
\text{v}_{k\uparrow}(\vec{r}) \\
\text{v}_{k\downarrow}(\vec{r})
\end{pmatrix}
= E_k
\begin{pmatrix}
\text{u}_{k\uparrow}(\vec{r}) \\
\text{u}_{k\downarrow}(\vec{r}) \\
\text{v}_{k\uparrow}(\vec{r}) \\
\text{v}_{k\downarrow}(\vec{r})
\end{pmatrix} ,
\end{gather}
\end{widetext}

\noindent
where $\mu$ is the chemical potential, and $E_k$ are the quasi-particle energies for each state. 
Because the particle number symmetry is broken in the presence  of the pairing correlations, the 
chemical potential is mathematically a Lagrange multiplier necessary to impose the 
additional condition of reproducing the correct average particle number. As is well established~\cite{Bulgac:1980,dobaczewski1984},
the $\text{v}$-components of the qpwfs have a finite norm if $E_k>0$, while the $\text{u}$-components belong to the continuum spectrum 
when $E_k-\mu>0$.
These equations describe 
a system with an even number of protons and neutrons. The generalization of these equations 
to odd numbers of protons or neutrons were discussed in Refs.~\cite{Ring:2004,Dobaczewski:1997} and recently
in Ref.~\cite{Bulgac:2020a}.

The local particle-hole Hamiltonian $h_{\sigma, \sigma'} (\vec{r})$ is obtained by
taking the appropriate functional derivatives of the energy density
functional. For the both Skyrme and SeaLL1 NEDFs  it takes the generic form 
\cite{Maruhn:2014,Ring:2004}:
\begin{align}\label{eq:spham} 
\begin{split}
h_q(\vec{r}) &= \left ( -\vec{\nabla} \cdot \frac{\hbar^2}{2m_q^*(\vec{r})}\vec{\nabla} + U_q(\vec{r}) \right ) 
- i \vec{W}_q(\vec{r}) \cdot (\vec{\nabla} \times \vec{\sigma})  \\
& +\vec{S}_q(\vec{r})\cdot\vec{\sigma} + \frac{1}{i}\left(\vec{\nabla}\cdot\vec{A}_q(\vec{r}) +\vec{A}_q(\vec{r})\cdot\vec{\nabla} \right),
\end{split}
\end{align}
where $q$ denotes neutron $n$ and proton $p$ channel.
The effective mass $m_q^*(\vec{r})$ is derived as 
\begin{align}
\frac{\hbar^2}{2m_q^*(\vec{r})}  = \frac{\delta \mathcal{E}}{\delta \tau_q(\vec{r})} = \frac{\hbar^2}{2m} + C^{\tau}_0\n_0 + \xi_q C^{\tau}_1 \n_1  \label{eq:mass_eff}
\end{align}
for Skyrme EDFs, where  $\xi_{n,p} = \pm 1$. In SeaLL1 NEDF, the coefficients $C_t^\tau$ are missing and $m^* = m$ is the bare nucleon mass.
The central-part of the mean-field potential $U(\vec{r})$ has the form
\begin{align}
U_q(\vec{r}) = \frac{\delta \mathcal{E}}{\delta \n_q(\vec{r})} & =  2C_0^{\n}\n_0 + 2\xi_q C_1^{\n}\n_1 +  + C_0^{\tau}\tau_0 + \xi_q C_1^{\tau}\tau_1 \nonumber \\
& + 2C_0^{\Delta\n}\nabla^2\n_0 + 2\xi_q C_1^{\Delta\n}\nabla^2\n_1 \nonumber\\
& + C_0^{\nabla J}\vec{\nabla}\cdot\vec{J}_0 + \xi_q C_1^{\nabla J}\vec{\nabla}\cdot\vec{J}_1 \nonumber \\
& + C_0^{\gamma}(\gamma+2)n_0^{\gamma+1} + 2\xi_q C_1^{\gamma}n_1 n_0^{\gamma} . \label{eq:u}
\end{align}
for Skyrme  NEDFs and 
\begin{align}
\begin{split}
U_q (\vec{r}) = & \frac{5}{3} a_0 \n_0^{2/3} + 2 b_0 \n_0 + \frac{7}{3} c_0 \n_0^{4/3} \\
& - \frac{a_1\n_1^2}{3  \n_0^{4/3}} + \frac{c_1 \n_1^2}{3 \n^{2/3}} \\
& - \frac{7a_2\n_1^4}{3 \n_0^{10/3}} - \frac{2 b_2 \n_1^4}{ \n_0^{3}} - \frac{5c_2 \n_1^4 }{ 3\n_0^{8/3} }\\
& + \xi_q \left( \frac{2 a_1 \n_1 }{ \n_0^{1/3}} + 2 b_1 \n_1 + 2 c_1 \n_1 \n^{1/3} + \right . \\
& \qquad  \left . \frac{4 a_2 \n_1^3}{ \n_0^{7/3} }+ \frac{4 b_2 \n_1^3 }{ \n_0^2} + \frac{4 c_2 \n_1^3 }{ \n_0^{5/3} }\right) \\
\end{split}
\end{align}
for SeaLL1 NEDF.

In both Skyrme and SeaLL1 NEDFs, the spin-orbit potential $\vec{W}_q(\vec{r})$ is given by 
\begin{align}\label{eq:spin_orbit_u}
\vec{W}_q(\vec{r}) = \frac{\delta \mathcal{E}}{\delta\vec{J}_q(\vec{r})} = C_0^{\nabla J}\grad{\n}_0 + \xi_q C_1^{\nabla J}\grad{\n}_1.
\end{align}
The applications of our code were so far restricted to even-even nuclei. In this case, 
only time-even terms in the NEDF contribute to the static solution. However, even 
if one considers only even-even nuclei, during the dynamical evolution the time-odd 
contributions from current and spin-densities need to be included in order to satisfy the Galilean invariance:
\begin{align}
\vec{S}_{q}(\vec{r}) & = C_0^{\nabla J} \vec{\nabla} \times \vec{j}_0 + C_1^{\nabla J} \vec{\nabla} \times \vec{j}_1 \label{eq:Sq}\\
\vec{A}_{q}(\vec{r}) & = C_0^{\tau} \vec{j}_0 + C_1^{\tau} \vec{j}_1 + \frac{1}{2} C_0^{\nabla J} \vec{\nabla} \times \vec{s}_0 + \frac{1}{2} \xi_q C_1^{\nabla J} \vec{\nabla} \times \vec{s}_1. \label{eq:Aq}
\end{align}
The local pairing field $\Delta_q (\vec{r})$ is
defined as a function of the anomalous density
\begin{align} \label{eq:pairingfield}
\Delta_q (\vec{r}) = - g_{\mathrm{eff}} (\vec{r}) \kappa_q (\vec{r}).
\end{align}

In SLDA both $\kappa(\vec{r})$ and $\Delta(\vec{r})$ have a local form, although 
in the original application of DFT to superconducting system \cite{Oliveira:1988}, 
a non-local pairing potential $\Delta(\vec{r}, \vec{r}')$ was used.
For a local pairing field, one can show that the anomalous density 
$\kappa(\vec{r},\vec{r}') = \sum_k v^*_{k\uparrow}(\vec{r}) \text{u}_{k\downarrow}(\vec{r}') 
\sim \frac{1}{\lvert \vec{r} - \vec{r}' \rvert}$ diverges for $\lvert \vec{r} - \vec{r}' \rvert \to 0$. 
 Also, in calculations 
with the Gogny interaction there is no divergence of the pairing field, due to the finite 
range of the interaction. The Gogny finite range interaction was introduced mainly 
to deal with this kind of divergence in a practical manner~\cite{Decharge:1980}, thus introducing 
into the phenomenology parameters, which have no microscopic meaning.
In the case of a local pairing field a simple renormalization of the pairing coupling constant to remove the 
divergence part in $\kappa(\vec{r})$ and in $\Delta(\vec{r})$  was suggested in Ref.~\cite{Bulgac:2002a}. For a lattice system 
with energy cutoff $E_c$, the effective pairing strength $g_{\mathrm{eff}}(\vec{r})$ in \cref{eq:pairing} is defined as
(for each neutron and proton)
\begin{subequations}\label{eq:geff}
\begin{align}
\frac{1}{g_{\mathrm{eff}}(\vec{r})} & = \left \{
\begin{array}{ll}
\frac{1}{g_0(\vec{r})} - \frac{m^*(\vec{r})k_c(\vec{r})}{2\pi^2\hbar^2} \left( 1 - \frac{k_F(\vec{r})}{2k_c(\vec{r})} \ln \frac{k_c(\vec{r})+k_F(\vec{r})}{k_c(\vec{r})-k_F(\vec{r})} \right), & k_F^2(\vec{r}) \geq 0 \\
\frac{1}{g_0(\vec{r})} - \frac{m^*(\vec{r}) k_c(\vec{r})}{2\pi^2\hbar^2}\left(1+ \frac{\abs{k_F(\vec{r})}}{k_c(\vec{r})} \arctan \frac{\abs{k_F(\vec{r})}}{k_c(\vec{r})} \right), & k_F^2(\vec{r})<0
\end{array}
\right. \\
E_c &= \frac{\hbar^2k_c^2(\vec{r})}{2m} + U(\vec{r}) - \mu, \\
\mu &= \frac{\hbar^2k_F^2(\vec{r})}{2m} + U(\vec{r}).
\end{align}
\end{subequations}
\cref{eq:geff} was derived using a spherical momentum space cutoff, as 
one does routinely in quantum field theory. When 
space is discretized on a lattice as in our approach, the energy cutoff 
$E_c$ should be smaller than the natural energy cutoff of the lattice 
$\frac{\hbar^2\pi^2}{2 m a^2}$ where $a$ is the lattice constant. In 
3D Cartesian coordinates with $dx=dy=dz=a$ the natural energy cutoff 
becomes $\frac{\hbar^2\pi^2}{2 m dx^2} + \frac{\hbar^2\pi^2}{2 m dy^2} + 
\frac{\hbar^2\pi^2}{2 m dz^2} = 3\frac{\hbar^2\pi^2}{2 m a^2}$ and for $E_c > \frac{\hbar^2\pi^2}{2 m a^2}$ 
\cref{eq:geff} is inapplicable. An expression of $g_\mathrm{eff}(\vec{r})$ 
in the case of a  3D lattice with the natural energy cutoff $3\frac{\hbar^2\pi^2}{2 m a^2}$ was 
suggested in Ref.~\cite{Castin:2004}
\begin{align}\label{eq:geff_3D}
\frac{1}{g_{\mathrm{eff}}(\vec{r})} = \frac{1}{g_0(\vec{r})} - \frac{m^*(\vec{r})}{4\pi^2\hbar^2}\frac{\pi}{dx} K, 
\end{align}
where $K$ is a numerical constant given by 
\begin{align*}
K = \frac{12}{\pi} \int^{\pi/4}_0 d\theta \ln(1+1/\cos^2\theta) = 2.442~749~607~806~335 \cdots.
\end{align*}

Performing fully self-consistent static SLDA calculations on a 3D lattice without any simplifications is
numerically expensive and we  typically resort to the following scheme. We first obtain fully self-consistent
solutions, often with help from N. Schunck (LLNL),  using the HFBHTO code~\cite{Perez:2017}. 
Those densities are converted from a basis of harmonic oscillator wave functions to the target 
3D spatial lattice. Since the size of the basis in 3D spatial lattice is significantly larger than the  
basis size used in HFBTHO calculations that affects the anomalous 
and kinetic energy densities, which converge very slowly with 
the energy cutoff we rerun the self-consistent calculations using the static SLDA code in order to 
determine the new neutron and proton chemical potentials only, assuming that the normal density 
is given correctly by the HFBHTO code. Typically only 3 self-consistent iterations with the 
static SLDA code are sufficient to achieve convergence.

\subsection{Time-dependent superfluid local density approximation (TDSLDA)}

The evolution of the qpwfs within the time-dependent superfluid local density approximation (TDSLDA) 
is described by the equations:

\begin{widetext}
\begin{align}\label{eq:tdslda}
i\hbar \frac{\partial \psi_k({\vec r},t) }{\partial t}=
i\hbar \frac{\partial}{\partial t}
\begin{pmatrix}
\text{u}_{k\uparrow} (\vec{r},t) \\
\text{u}_{k\downarrow}(\vec{r},t) \\
\text{v}_{k\uparrow}(\vec{r},t) \\
\text{v}_{k\downarrow}(\vec{r},t)
\end{pmatrix}
= {\cal H}(\vec{r},t) 
\begin{pmatrix}
\text{u}_{k\uparrow} (\vec{r},t) \\
\text{u}_{k\downarrow}(\vec{r},t) \\
\text{v}_{k\uparrow}(\vec{r},t) \\
\text{v}_{k\downarrow}(\vec{r},t)
\end{pmatrix}
=
\begin{pmatrix}
h_{\uparrow \uparrow}(\vec{r},t) -\mu  & h_{\uparrow \downarrow}(\vec{r},t) & 0 & \Delta(\vec{r},t) \\
h_{\downarrow \uparrow}(\vec{r},t) & h_{\downarrow \downarrow}(\vec{r},t) - \mu& -\Delta(\vec{r},t) & 0 \\
0 & -\Delta^*(\vec{r},t) &  -h^*_{\uparrow \uparrow}(\vec{r},t) +\mu & -h^*_{\uparrow \downarrow}(\vec{r},t) \\
\Delta^*(\vec{r},t) & 0 & -h^*_{\downarrow \uparrow}(\vec{r},t) & -h^*_{\downarrow \downarrow}(\vec{r},t) + \mu
\end{pmatrix}
\begin{pmatrix}
\text{u}_{k\uparrow}(\vec{r},t) \\
\text{u}_{k\downarrow}(\vec{r},t) \\
\text{v}_{k\uparrow}(\vec{r},t) \\
\text{v}_{k\downarrow}(\vec{r},t)
\end{pmatrix}. 
\end{align} 
\end{widetext}
During the evolution, the solution remains a single generalized Slater 
determinant of time-dependent qpwfs, even in the case of fission, where 
separation between fragments can be observed in the densities of the fragments.  Note, 
that in the time dependent equations the chemical potential can be dropped, as it can be 
removed with a trivial gauge transformation. Moreover, unlike the static SLDA 
\cref{eq:hfb}, the \cref{eq:tdslda} has the same form for even-even, odd, and 
odd-odd nuclei, if the chemical potential $\mu$ is dropped.

\subsubsection{External boosts and external potentials}
In studies of nuclear reactions and giant resonances, the nucleus is ``boosted'' 
at the beginning of the evolution. Such boost is realized by 
performing a gauge transformation on each qpwf as 
\begin{align}\label{eq:kick}
\begin{pmatrix}
\text{u}_{k\sigma}(\vec{r},t) \\
\text{v}_{k\sigma}(\vec{r},t)
\end{pmatrix}
\to
\begin{pmatrix}
e^{i\chi({\vec r})} & 0 \\
0 & e^{-i\chi({\vec r})}
\end{pmatrix}
\begin{pmatrix}
\text{u}_{k\sigma}(\vec{r},t) \\
\text{v}_{k\sigma}(\vec{r},t)
\end{pmatrix},
\end{align}
where $\sigma=\uparrow,\downarrow$.
The spatial profile $\chi(\vec{r})$ should be chosen appropriately 
for different situations. For example. when $\chi(\vec{r}) = \vec{p} \cdot \vec{r} / \hbar$,  
the nucleus will gain an initial velocity $\hbar \vec{\nabla} \chi / m = \vec{p}/m$ 
with $m$ the average nucleon mass. 

In addition, during the evolution, the system can also be coupled to 
external time dependent scalar and vector fields~\cite{Stetcu:2015,Stetcu:2011,Bulgac:2019b,Bulgac:2019a}
\begin{align}
\!\!\!
\hat{h}_q \to \hat{h}_q + U_{q,\mathrm{ext}}^0 (\vec{r},t) 
- \frac{1}{2} \left [\vec{U}_{q,\mathrm{ext}} (\vec{r},t)\cdot \hat{\vec{p}} 
+ \hat{\vec{p}} \cdot \vec{U}_{q,\mathrm{ext}}(\vec{r},t)\right ].
\end{align}
Different choices for the external potentials allow one to study 
the response of the nuclear system to various kind of probes.

\subsubsection{Center of mass motion and rotation}

The presence of external potentials or boosts sometimes lead to center 
of mass motions and rotations of the system, see e.g. Ref.~\cite{Stetcu:2015}. In order to follow the internal 
motion of nucleus in the moving and/or in the  rotating frame, we need to introduce 
extra terms in the Hamiltonian to counter balance these collective motions \cite{Stetcu:2015}. 
For the center of mass motion, we perform a a transformation for each qpwf as 
\begin{align}\label{eq:translation1}
\begin{pmatrix}
\text{u}_{k\sigma}(\vec{r},t) \\
\text{v}_{k\sigma}(\vec{r},t)
\end{pmatrix}
\to
\begin{pmatrix}
  \exp \left[ \frac{i \vec{R}(t) \cdot \hat{\vec{p}}}{\hbar} \right] & 0 \\
0 &  \exp \left[ -\frac{i \vec{R}(t) \cdot \hat{\vec{p}}}{\hbar} \right]
\end{pmatrix}
\begin{pmatrix}
\text{u}_{k\sigma}(\vec{r},t) \\
\text{v}_{k\sigma}(\vec{r},t)
\end{pmatrix},
\end{align}
where $\vec{R}(t)$ describes the center of mass motion and $\hat{\vec{p}}$ is the momentum operator. Then the 
single-particle Hamiltonians are replaced as follows in Eqs.~\cref{eq:tdslda}
\begin{align} \label{eq:tdsh_cm}
h_{\sigma \sigma}(\vec{r},t) \to h_{\sigma \sigma}(\vec{r},t)- \vec{\text v}_{\mathrm{cm}}(t) \cdot \hat{\vec{p}}
\end{align}
where $\vec{\text v}_{\mathrm{cm}}(t) = \dot{\vec{R}}(t)$ is the velocity of the c.m. motion and $\sigma =\uparrow,\downarrow$. 
In practice this is calculated as 
\begin{align}\label{eq:vel}
\vec{\text v}_{\mathrm{cm}}(t) = \int d^3\!r \frac{\hbar \vec{j}(\vec{r},t) }{M},
\end{align}
where $\vec{j}(\vec{r},t) = \vec{j}_n(\vec{r},t) + \vec{j}_p(\vec{r},t)$ is the total current 
density, see \cref{eq:j},  and $M$ is the total mass of the nucleus.

Similar to the center of mass motion, the rotation of the system can 
also be balanced by making a transformation on the qpwfs like 
\begin{align}\label{eq:translation2}
\begin{pmatrix}
\text{u}_{k\sigma}(\vec{r},t) \\
\text{v}_{k\sigma}(\vec{r},t)
\end{pmatrix}
\to
\begin{pmatrix}
 \exp \left[ \frac{i \vec{\theta}(t) \cdot \hat{\vec{j}}}{\hbar} \right]  & 0 \\
0 & \exp \left[ -\frac{i \vec{\theta}(t) \cdot \hat{\vec{j}}}{\hbar} \right] 
\end{pmatrix}
\begin{pmatrix}
\text{u}_{k\sigma}(\vec{r},t) \\
\text{v}_{k\sigma}(\vec{r},t)
\end{pmatrix},
\end{align}
where $\vec{\theta}(t) = \theta_0(t) \hat{\theta}(t)$ is the vector of rotation angle 
for the system away from the $z$-axis and $\hat{\vec{j}}$ is the single-particle angular momentum operator
\begin{align}
\hat{\vec{j}}= \hat{\vec{l}} + \hat{\vec{s}}, \quad \hat{\vec{l}} = \hat{\vec{r}} \times \hat{\vec{p}}.
\end{align}
 $\vec{\theta}$ is determined according to the following procedure. 
 We calculate the instantaneous mass quadrupole matrix of the system as 
\begin{align}\label{eq:qua_matrix}
\quad Q_{ij} = \int d^3\!r \vec{r}_i \vec{r}_j \n(\vec{r}), \quad i, j = x,y,z
\end{align}
and diagonalizing $Q$ we obtain the rotation matrix $R$:
\begin{align}
Q = R D R^T
\end{align}
which has a structure 
\begin{align}
R_{ik} = \cos \theta_0 \delta_{ik} + (1-\cos \theta_0) n_i n_k - \sin\theta_0 \varepsilon_{ikl}n_l
\end{align}
where $(n_x, n_y, n_z)$ is the unit vector of rotation axis $\hat{\theta}$. These 
quantities can be determined from the following relations
\begin{align}
\mathrm{Tr}~R = 1 + 2 \cos \theta_0, \\
 R - R^T = 2\sin \theta_0
\begin{pmatrix}
0 & -n_z & n_y \\
n_z & 0 & -n_x \\
-n_y & n_x & 0
\end{pmatrix}
\end{align}
In practice, we usually ignore the rotation along $z$ axis, then the rotation angle can be determined simply as 
\begin{align}
\vec{\theta} = \vec{R}_3 \times \hat{z}
\end{align}
where $\vec{R}_3=(n_y,-n_x,0)$ is the third column of $R$ matrix and $\hat{z} = (0,0,1)$ is 
the unit vector of $z$-direction. The single-particle Hamiltonian changes accordingly as follows
\begin{align}\label{eq:tdsh_rot}
h_{\sigma \sigma}(\vec{r},t) \to h_{\sigma \sigma}(\vec{r},t)-  \vec{\omega}(t) \cdot \hat{\vec{j}}
\end{align}
where $\vec{\omega}(t) = \dot{\vec{\theta}}(t)$ is the angular velocity. 
Using this particular transformation to a rotating frame reduces considerably the complexity 
of a nucleus-nucleus simulation at a finite impact parameter. During such a collision the long 
side of the simulation box is always aligned with the line joining the two reaction partners.
The transformation back to the the laboratory frame is straightforward.

\subsubsection{Miscellaneous quantities}

Various (time-dependent) quantities are also calculated in the code, such as:
\begin{itemize}
\item
the center of mass (for nucleons, protons, and neutrons)
\begin{align}\label{eq:center_of_mass}
\vec{r}_{\mathrm{cm}} = (x_{\mathrm{cm}},y_{\mathrm{cm}},z_{\mathrm{cm}})
= \frac{\int d^3\!r \vec{r} n(\vec{r})}{\int d^3\!r n(\vec{r})},
\end{align}
\item 
the deformation parameters
\begin{subequations} \label{eq:deform}
\begin{align}
&Q_{lm} = \int d^3\!r \hat{Q}_{lm} \n(\vec{r}), \quad \hat{Q}_{lm} \propto r^l Y_{lm} \label{eq:multiple}
\end{align}
where $Y_{lm}$ is the spherical harmonics function, and
\begin{align}
&\hat{Q}_{20} = (2z'^2 - x'^2 - y'^2), \label{eq:Q20}\\
&\hat{Q}_{30} = z' (2z'^2-3x'^2-3y'^2), \label{eq:Q30} \\
&\hat{Q}_{40} = 35z'^4 - 30z'^2r'^2 + 3r'^4, \label{eq:Q40}\\
&\hat{Q}_{21} = (x+iy)z, \quad \hat{Q}_{22} = (x+iy)^2, \label{eq:Q21_Q22}
\end{align}
\end{subequations}
with $x' = x - x_{\mathrm{cm}}, y' = y - y_{\mathrm{cm}}, z' = z - z_{\mathrm{cm}}$ and $r'^2 = x'^2 + y'^2 + z'^2$.
\item 
the average pairing gap:
\begin{align}
\langle \Delta_q \rangle = \frac{\int d^3\!r \lvert \Delta_q(\vec{r}) \rvert \n_q(\vec{r}) }{\int d^3\!r \n_q(\vec{r})},
\end{align}
\item
the collective flow energy
\begin{align}
E_{\mathrm{coll}} = \sum_q \int d^3\!r \frac{\hbar^2}{2m} \vec{j}_q^2(\vec{r}),
\end{align}
\item
the center of mass kinetic energy
\begin{align}
E_{\mathrm{cm}} = \frac{1}{2} M \vec{\text{v}}^2_{\mathrm{cm}},
\end{align}
\item 
the total kinetic energy (TKE)
\begin{subequations}
\begin{align}
\text{TKE} = \frac{1}{2}mA_{\mathrm{H}} \vec{\text v}_{\mathrm{H}}^2 + \frac{1}{2}mA_{\mathrm{L}} \vec{\text v}_{\mathrm{L}}^2 + E_{\mathrm{Coul}}, \label{eq:TKE}
\end{align}
with the velocity of the fragment $f = H, L$ given by 
\begin{align}
\vec{\text v}_f = \frac{1}{mA_f} \int_{V_f} d^3\!r \vec{j}(\vec{r}), \quad A_f = \int_{V_f} d^3\!r n(\vec{r}),
\end{align}
\item
where $\vec{j}(\vec{r})$ and $\n(\vec{r})$ are the total current densities and number densities respectively, 
and  the integral is performed over the appropriate half-box $V_f$ where each fragment is located. 
\item The Coulomb interaction energy between fragments (direct term only) is given by 
\begin{align}
E_{\mathrm{Coul}} = e^2 \int_{V_{\mathrm{H}}} d^3\!r_1\int_{V_{\mathrm{L}}} d^3\!r_2 \frac{n_p(\vec{r}_1)n_p(\vec{r}_2)}{\lvert \vec{r}_1 - \vec{r}_2 \rvert} ,\label{eq:ecoul_between}
\end{align}
where $n_p(\vec{r})$ is the proton number density.
\end{subequations}
\end{itemize}

\section{Numerical Implementation}

The first version of the code was developed by AB and Yongle Yu in 2007 for cold Fermi gases and written in Fortran for sequential execution, 
and was ported to Jaguar at OLCF Oak Ridge by KJR, who implemented the parallelization of the code in both Fortran and C~\cite{Bulgac:2008x}. 
Yuan-Lung Luo wrote the first version of the C code and used the parallel version for the
simulation of the creation, dynamics, crossing, and recombination of quantized vortices in the 
unitary Fermi gas~\cite{Bulgac:2011}. 
The various versions of the code were subsequently used on in other 
studies~\cite{Stetcu:2011,Bulgac:2012,bulgac2012,bulgac2013a,Bulgac:2013}.
Gabriel Wlaz{\l}owski created  a CUDA set of complex arithmetic functions, which 
was used in the first hybrid CPU-GPU version of the code on Titan at OFCL ORNL in 2013, to simulate
the nascency and evolution of quantized vortex rings in a cold atom system~\cite{Bulgac:2014}. 
IS ported the C version of the code to nuclear systems, and together with KJR 
ported the nuclear code to use both MPI and CUDA for parallel execution targeting Titan at OLCF Oak Ridge - a hybrid computer system composed of CPUs and Nvidia GPUs. 
That type of  code was used for all subsequent studies on hybrid computer systems in 
Refs.\cite{Stetcu:2015,Bulgac:2016,Bulgac:2016x,Bulgac:2017,Bulgac:2019b,Bulgac:2019c,Bulgac:2020,Wlazlowski:2016,Wlazlowski:2016}
on Titan and Summit at OLCF Oak Ridge, on
Piz Daint at CSCS, Lugano, on Tsubame at Tokyo Technological Institute, on
Kodiak at LANL, Los Alamos, and on Lassen and Sierra at LLNL, 
Livermore. 
While the numerical implementation has stayed basically unchanged during these years, 
only with relatively minor improvements of the numerical details,
the computer implementation evolved accordingly with the evolution of the computer hardware.  

\subsection{Discretization} \label{sec:meshgrid}
In both the static and the dynamic codes, all the spatial functions are 
discretized on a 3D Cartesian lattice of size $N_x, N_y, N_z$ with lattice 
constant $dx=dy=dz=a$. In each direction of length $L=Na$, the spatial 
coordinate $x_n$ and Fourier component $k_n$ are discretized as follows:
\begin{align}
x_n & = n a, \quad n = 0, \cdots, N-1 \\
k_n & = 
\left \{
\begin{array}{cc}
2n\pi/L, & n = 0, \cdots, N/2-1 \\
(2n-N)\pi/L, & n = N/2, \cdots, N-1
\end{array}
\right. ,
\end{align}
consistent with boundary conditions. The apparent absence of reflection symmetry 
$x\rightarrow -x$ has no practical consequence, since all functions 
satisfy periodic boundary conditions 
$F(x_0)=F(x_{N})$. 
In our discretized simulation box the products of plane waves $\exp(ik_nx_m)$ with $n,m=0,N-1$ in
all 3 directions form a full set~\cite{Bulgac:2013}, see below also. 
In calculations of any observables which depend explicitly on 
coordinates we always use $\vec{r} \rightarrow \vec{r}-\vec{R}_\textrm{cm}$, and  
in case of fragments we use the center of mass of the fragments respectively, as naturally the 
arbitrary choice of the origin of the coordinate system should not play a role.

When discretized on the (3D) lattice, both wavefunctions and the Hamiltonian 
are represented in the discrete variable representation (DVR) \cite{Bulgac:2013}, 
which in numerical analysis is sometimes referred to as the 
Lagrange-mesh method \cite{Baye:2015}. In one dimension, the functions are represented as
\begin{align}\label{eq:function_dvr}
\psi(x) = \sum_k a \psi(x_k) f_k(x)
\end{align}
where the DVR basis states $f_k$s form a complete orthonormal set on the special lattice
\begin{subequations}\label{eq:dvr_basis}
\begin{align}
\begin{split}
f_k(x_l) &= \sum_{n=-N/2}^{N/2-1} \frac{1}{L} \exp[ik_n(x_l-x_k)]  \\
& = \left \{ 
\begin{array}{ll}
\frac{\sin \pi(k-l)}{Na} \cot \frac{\pi(k-l)}{N} = 0 & k\neq l,\\
1/a & k = l,
\end{array}
\right .
\end{split}
\end{align}
and 
\begin{align}
\quad \inner{f_k}{f_l} = \sum_n a f_k(x_n) f_l(x_n) = \delta_{kl}.
\end{align}
\end{subequations}
Once the length of the simulation box and the momentum cutoff $p_\text{cut}=\pi\hbar/a$ is chosen
the DVR basis is numerically the minimal basis required to solve the Schr{\"o}diger equation~\cite{Bulgac:2013x}.
If the length of the simulation box $L$ and the lattice constant $a$ are chosen 
appropriately on physical arguments, any further increase in $L$ and decrease in $a$ will lead only 
to exponentially small corrections, see Ref. \cite{Bulgac:2013x} and references therein.

In static SLDA calculations, one needs to diagonalize Hamiltonian matrix 
in \cref{eq:hfb} in a series of iterations, until self consistent solutions are found. 
All the local potential matrix elements, which depend in general on densities 
and current densities, have a simple diagonal representation
\begin{align}
\langle f_k | U | f_l \rangle = \sum a f^*_k(x_n) U(x_n) f_l(x_n) = U(x_k) \delta_{kl}.
\end{align}
The non-diagonal matrix elements source from the spatial first $\partial_x$ 
and second derivative $\partial_{xx}$ in the kinetic and spin-orbit 
terms, which in DVR basis can be represented as  
\begin{subequations}
\begin{align}\label{eq:d1}
\!\!\!\!\!
(\partial_x)_{nm} = \frac{\pi}{Na} (-1)^{n-m} (1-\delta_{nm}) \cot \left( \frac{\pi(n-m)}{N}\right) , 
\end{align}
\begin{align}\label{eq:d2}
\!\!\!\!
(\partial_{xx})_{nm} = \frac{\pi^2}{2N^2a^2} \frac{(-1)^{(n-m)} (\delta_{nm}-1)}{\sin^2 \frac{\pi(n-m)}{N}} 
- \frac{\pi^2}{3a^2}\left( 1+ \frac{2}{N^2}\right) \delta_{nm}, 
\end{align}
\end{subequations}
see Refs.~\cite{Bulgac:2013x,Maruhn:2014} for details. In 3D this DVR representation of the Hamiltonian leads 
to a very sparse matrix with only $4N_xN_yN_z-6$ non-vanishing matrix elements of the 
$4N_xN_yN_z\times 4N_xN_yN_z$ discretized Hamiltonian matrix. 

In the time-dependent calculation, the Hamiltonian matrix is not explicitly 
evaluated. Instead,  we directly compute the matrix-vector products $H\psi$ (twice) 
in each time-step. Besides the trivial products of the local potentials and 
wave functions, in the application of the Hamiltonian on an arbitrary state, 
the most time-consuming part is the calculations of gradient and Laplacian 
of the wave functions. In order to make this operation efficient 
we use the fast Fourier transforms (FFT) 
\begin{subequations}
\begin{align}
\mathcal{F}[f(x)]&: \tilde{f}(k_n) = \sum_{l=0}^{N-1} \exp (-i k_n x_l) f(x_l), \\
\mathcal{F}^{-1}[\tilde{f}(k)]&: f(x_l) = \frac{1}{N}\sum_{n=0}^{N-1} \exp (i k_n x_l) \tilde{f}(k_n),
\end{align}
\end{subequations}
with $\mathcal{F}$ and $\mathcal{F}^{-1}$ the direct and inverse 
Fourier transforms. Hence, the application of the derivative operator $m$ times is efficiently evaluated as follows: 
\begin{align}
\mathcal{D}^{(m)} f(x) = \mathcal{F}^{-1} \left[ (ik)^{m} \tilde{f}(k) \right].
\end{align}
For odd derivatives there is a loss of symmetry that can induce small 
imaginary part into derivatives that should be purely real. In order to 
avoid issues, we set $\tilde{f}(-k_{N/2}) = 0$~\cite{Maruhn:2014,Bulgac:2013,trefethen2000spectral}.
Since an FFT requires ${\cal O}(N\ln N)$ of floating point operations (FLOPs) the approach of calculating
derivatives is competitive and often faster than any accurate finite difference formula 
and moreover exact to machine precision. 

For better numerical accuracy, we avoid computing first-order derivatives if possible
 and we take advantages of standard relationships such as
\begin{align}
& \vec{\nabla} F({\vec r})\cdot \vec{\nabla} G({\vec r}) \rightarrow \\
& \frac{1}{2}
\Big( 
\Delta [ F({\vec r}) G({\vec r}) ] - \Delta [F({\vec r})] G({\vec r})  -F({\vec r})\Delta [G({\vec r}) ]
\Big), \nonumber
\end{align}
for increased numerical accuracy. The evaluation of first order derivatives requires the elimination 
of the highest frequency in the Fourier transform for numerical accuracy. If  couplings to gauge fields is 
required, as in Ref.~\cite{Stetcu:2015,Stetcu:2011}, or when evaluating terms linear in momentum,
we use the discretized symmetrized form
\begin{align}
& \vec{A}({r},t)\cdot \vec{\nabla}\psi({\vec{r},t}) \rightarrow  \\ 
& \frac{1}{2}
\left [ \vec{A}({r},t)\cdot \vec{\nabla}\psi({\vec{r},t}) 
     + \vec{\nabla}\psi({\vec{r},t}) \cdot   \vec{A}({r},t) \right  ]. \nonumber
\end{align}
When evaluating first order derivatives of products of functions we use Leibniz rule 
\begin{align}
\vec{\nabla} \left [A(\vec{r})B(\vec{r})\right ] \rightarrow \
 B(\vec{r})  \vec{\nabla} A(\vec{r})+  A(\vec{r}) \vec{\nabla} B(\vec{r}).
 \end{align}
The use of this rule is particularly important to ensure numerically accurate gauge invariance.
As discussed in Refs. \cite{Bulgac:2013,Ryssens:2015} with a 
careful choice of the size of the box and of the spatial lattice constant one can achieve very high 
numerical accuracy with relatively large values of lattice constant $l$. For example, 
the kinetic energy term in \cref{eq:spham} can be rewritten as 
\begin{align}
\begin{split}
& - \vec{\nabla} \cdot \frac{\hbar^2}{2m^*(\vec{r})}\vec{\nabla} {\text v}_{k\sigma}(\vec{r})  
 = -\frac{1}{2}   \frac{\hbar^2}{2m^*(\vec{r})} \nabla^2 {\text v}_{k\sigma}(\vec{r}) \\
& +\frac{1}{2}\nabla^2 \left( \frac{\hbar^2}{2m^*(\vec{r})}{\text v}_{k\sigma}(\vec{r}) \right) 
 -\frac{1}{2}\left( \nabla^2 \frac{\hbar^2}{2m^*(\vec{r})} \right) {\text v}_{k\sigma}(\vec{r}),
\end{split}
\end{align}
which corresponds to a $T$ symmetric matrix representation in the static solver as 
\begin{align}\label{eq:matrix_T}
\begin{split}
T_{nm} = \left( - \vec{\nabla} \cdot \frac{\hbar^2}{2m^*(\vec{r})}\vec{\nabla} \right)_{nm} 
 = & -\frac{1}{2} (\nabla^2)_{nm} \left(\frac{\hbar^2}{2m^*_n} + \frac{\hbar^2}{2m^*_m} \right) \\
& + \frac{1}{2} \left( \nabla^2 \frac{\hbar^2}{2m^*} \right)_n \delta_{nm}
\end{split}
\end{align}
where the laplacian operator $\nabla^2$ is a symmetric matrix as in \cref{eq:d2}.

In practice, we also find that better accuracy can be achieved 
when a special symmetrization is  performed for the spin-orbit term
\begin{align}
\frac{1}{i} \vec{W} (\vec{r}) \cdot \left( \vec{\nabla} \times \vec{\sigma} \right)\psi = 
\frac{1}{2i} \left[ \vec{W} (\vec{r}) \cdot \left( \vec{\nabla} \times \vec{\sigma} \right)\psi 
+ \vec{\nabla} \cdot \left( \vec{\sigma} \times \vec{W} (\vec{r}) \psi\right) \right].
\end{align}

\subsection{The Coulomb potential}

In computing the Coulomb 
potential $V_c(\vec{r})$ generated by the 
charge(proton) density $n_p(\vec{r})$, 
we use the method described in Ref. \cite{Castro:2003} to solve the Poisson equation 
in order to eliminate the contributions 
from images, which  are inherent when using periodic boundary conditions:
\begin{subequations}
\begin{align}
\vec{\nabla}^2 \Phi(\vec{r}) & = 4\pi e^2 n(\vec{r}), \\
\Phi(\vec{r}) & = \int d^3 \vec{r}' \frac{e^2 n(\vec{r})}{\abs{\vec{r} - \vec{r}'}}. \label{eq:poisson}
\end{align}
\end{subequations}
In numerical implementation \cite{Stetcu:2015,Bulgac:2016,magierski2017,jin2017,Bulgac:2017} 
the convolution \cref{eq:poisson} is solved by {FFT} $\mathcal{F}$ via
\begin{align}\label{eq:vc_fft}
& \Phi(\vec{r}) = \int \frac{d^3k}{(2\pi)^3} \frac{4\pi e^2 \tilde{n}(\vec{k})}{k^2} 
\exp(i\vec{k}\cdot\vec{r}) = e^2\mathcal{F}^{-1} [\tilde{n}(\vec{k})\tilde{f}(\vec{k})], \\
& \tilde{n}(\vec{k}) = \mathcal{F}[n(\vec{r})]
\end{align}
To avoid the infrared divergence for $\vec{k}\to 0$ we use a truncated 
kernel $\tilde{f}(k)$ \cite{Castro:2003}, considering a modified Coulomb potential
\begin{align}\label{eq:fk}
& f(r) = 
\left \{
\begin{array}{ll}
1/r, & \text{for}~r < D\\
0, & \text{otherwise}
\end{array}
\right .
= \int \frac{d^3k}{(2\pi)^3} \tilde{f}(\vec{k}),\\
& \tilde{f}(\vec{k}) = \left \{
\begin{array}{ll}
4\pi e^2\left[1-\cos (\abs{\vec{k}}D) \right]/k^2, & \lvert \vec{k} \rvert \neq 0 \\
2\pi e^2 D^2, & \lvert \vec{k} \rvert = 0
\end{array}
\right. 
\end{align} 
If in a cubic box $L_x=L_y=L_z=L$ we choose $D = \sqrt{3}L$ we 
can compute the Coulomb potential due to the charge distribution 
inside the box, thus eliminating any contribution from the neighboring cells. 
In a rectangular 3D box, we select $L$ to be the longest dimension 
among $L_x, L_y, L_z$ and the \cref{eq:vc_fft} is realized by the summation 
\begin{align}\label{eq:vc3}
\Phi(\vec{r}) = \frac{1}{27N^3} \sum_{\vec{k} \in (3L)^3} e^2 \tilde{n}(\vec{k}) \tilde{f}(\vec{k}) \exp(i\vec{k}\cdot\vec{r}),
\end{align}
with a number of floating point operations is $\sim 27N^3\ln (27N^3)$ 
where $N={\text{max}}(N_x,N_y,N_z)$. 

Because of the summation over $27N^3$ points, in the case when one 
dimension is much larger than the other two, as we chose often in the 
case of fission dynamics, the calculation becomes considerably more expensive. 
In such a case, the computational cost can be reduced by the following method.  
Consider a function $f(x)$ on the interval $(0,L)$ which we extend to the 
interval $(-L,2L)$ by adding zeros outside the main interval and the new function called $g(x)$
\begin{align}\label{eq:gx}
g(x) = 
\left \{ 
\begin{array}{ll}
f(x), & \text{if}~0 \leq x \leq L \\
0, & \text{otherwise}.
\end{array}
\right . 
\end{align}

Apart from normalization one can define two Fourier transforms, one 
on the interval $(0,L)$ and the other on the interval $(-L,2L)$, discretized 
with the same lattice constant $\Delta x$. The discrete Fourier transform 
on the interval $(0,L)$ will have Fourier components at momenta
\begin{align}
k_i=-\frac{\pi}{\Delta x}+\frac{2\pi i}{L} \mbox{ for } i=0, \dots N-1,
\end{align}
while the other one will have components at momenta
\begin{align}
\tilde k_i=-\frac{\pi}{\Delta x}+\frac{2\pi i}{3L} \mbox{ for } i=0, \dots 3N-1,
\end{align}
hence $\tilde k_{3i}=k_i$, for $=0, \dots N-1$. Indeed, $N$ of the momenta 
on the $3N$ discretization are the same with the momenta on the $N$ 
discretization, the other $2N$ momenta being defined as $k_i+2\pi/(3L)$ 
and $k_i+4\pi/(3L)$, with $i=0,\dots N-1$. For the arbitrary function $f(x)$ non-vanishing on the interval $(0,L)$ 
we can perform the following three Fourier transforms:
\begin{subequations}
\begin{align}
h_n &= \sum_l f(x_l) \exp(-ik_n x_l) \\
h_{n+1/3} &= \sum_l f(x_l) \exp\left( -i \frac{2\pi}{3L}x_l \right) \exp(-ik_n x_l) \\
h_{n+2/3} &= \sum_l f(x_l) \exp\left( -i \frac{4\pi}{3L}x_l \right) \exp(-ik_n x_l).
\end{align}
\end{subequations}
In terms of $h_n$, $h_{n+1/3}$, and $h_{n+2/3}$, one can prove the 
following relationship for  function $g(x)$ defined in Eq. (\ref{eq:gx})
\begin{align}\label{eq:gx_fft}
\begin{split}
g(x_l) &= \frac{1}{3}\sum_n\exp(ik_nx_l) \left [h_n + \exp\left(i\frac{2\pi}{3L}x_l\right) h_{n+1/3}\right. \\
& +\left. \exp\left(i\frac{4\pi}{3L}x_l\right)h_{n+2/3} \right ] \exp(ik_nx_l) ,
\end{split}
\end{align}
where the 1/3 factor comes from the different normalizations of the 
discrete Fourier transforms using discretizations with $N$ and $3N$ 

Thus one reduces considerably, as one needs three forward 
and three backward Fourier transforms in 1D  in the interval $(0,L)$ and 
the number of operations reduces to $\order{6N \ln N}$. 
In 3D one has to perform 27 forward and 27 backward Fourier 
transforms  in a $L^3$ simulation box only. Moreover, all these
Fourier transforms can be performed in parallel. The complete form 
of such a decomposition of \cref{eq:vc3} is 
\begin{widetext}
\begin{subequations}
\begin{align}
\begin{split}
\Phi(\vec{r}) =  \frac{1}{27N^3} \sum_{\xi,\eta,\zeta=0}^2 \left[ \sum_{k\in L^3} e^2 
\tilde{n}_{klm}(\vec{k}) \tilde{f}\left( \vec{k} + \left( \xi\frac{2\pi}{3L}, \eta\frac{2\pi}{3L}, \zeta\frac{2\pi}{3L} \right) \right) \exp(i\vec{k}\cdot\vec{r}) \right] 
 \times \exp \left(i \left(\xi\frac{2\pi}{3L}x + \eta\frac{2\pi}{3L}y + \zeta\frac{2\pi}{3L}z \right) \right)
\end{split}
\end{align}
where 
\begin{align}
\tilde{n}_{klm}(\vec{k}) = \sum_{\vec{r}\in L^3} n(x,y,z) \exp\left( -i \left( \xi\frac{2\pi}{L}x 
+ \eta\frac{2\pi}{L}y + \zeta\frac{2\pi}{L}z \right) \right) \exp(-i\vec{k}\cdot \vec{r}).
\end{align}
\end{subequations}
\end{widetext}

\subsection{Static SLDA }
\subsubsection{Self-consistent iterations}
The static solution is obtained by solving the SLDA equation \cref{eq:hfb} self-consistently. The static solver starts from a set of initial local densities denoted generically by $\vect{n}^{(0)}$. Two options are provided for starting the calculations:
\begin{itemize}
\item the initial guess of local densities $\vect{n}^{(0)}$ are constructed from periodic 3D Gaussian-like functions with appropriate saturation and surface properties for the desired nucleus;
\item the initial densities $\vect{n}^{(0)}$ are read from the disk; in this case, previously calculated densities with this solver or generated with other solvers can be used to continue the calculation on the lattice, or to produce the initial conditions by performing one diagonalization. Sometimes one needs to perform several iterations however. For example, if the densities originated from a different solver, the treatment of pairing might be different. In the time-dependent code we use sometimes a spherical momentum cutoff and sometimes a cubic momentum cutoff. In such cases one needs to determine the chemical potential which corresponds to the correct particle number. Typically three iterations are sufficient.  The first iteration with the original chemical potential leads to an incorrect particle number.  One then uses the procedure described below. In other cases the original densities for two nuclei in a collision simulation are obtained for isolated nuclei. When placed in a simulation box at a finite distance the long ranged Coulomb potential of one nucleus affects the proton chemical potential of the other nucleus. If the nuclei are sufficiently far apart it is sufficient to correct the chemical potential of each nucleus by the Coulomb  field created by the other nucleus considered as point charge.    If the nuclei are different their chemical potentials are different and in this case one has to interpolate them from one value to the other using a smooth interpolating function. In the time-dependent code the chemmical potential does not need to appear and it can be removed like any other constraint use in static calculations. 
\end{itemize}
 The various local potentials $\vect{V}^{(0)}$ and the resulting Hamiltonian matrix are generated using densities $\vect{n}^{(0)}$. In the self-consistent iteration $m$, the qpwfs $\psi_k^{(m)}$ and their corresponding quasiparticle energies $E_k^{(m)}$ are obtained by a direct diagonalization of the Hamiltonian using  the \texttt{pzheevd()} function included in the ScaLAPACK library. Next, the new local densities are constructed from the qpwfs using \cref{eq:local_densities} and the new potentials $\vect{V}^{\mathrm{new}}$ are generated as well. The chemical potential $\mu$ should also be updated for the convergence of the particle number. In this code we use the relation between the change in chemical potential and the change of particle number from the Thomas-Fermi (TF) approximation (separately for neutrons and protons)
\begin{align}
\delta\mu = \frac{2}{3}E_F \frac{\delta N}{N}
\end{align}
where $E_F = \hbar^2 k_F^2 / 2m \approx 35~\mathrm{MeV}$ is the Fermi energy of the infinite symmetric nuclear matter. Thus the new chemical potential is updated like 
\begin{align}
\mu^{\mathrm{new}} = \mu^{(m-1)} - \frac{2}{3}E_F (N^{(m)} - N_0) /N_0,
\end{align}
where $N_0$ is the desired particle number.

\begin{figure}[htbp]
\includegraphics[scale=0.5]{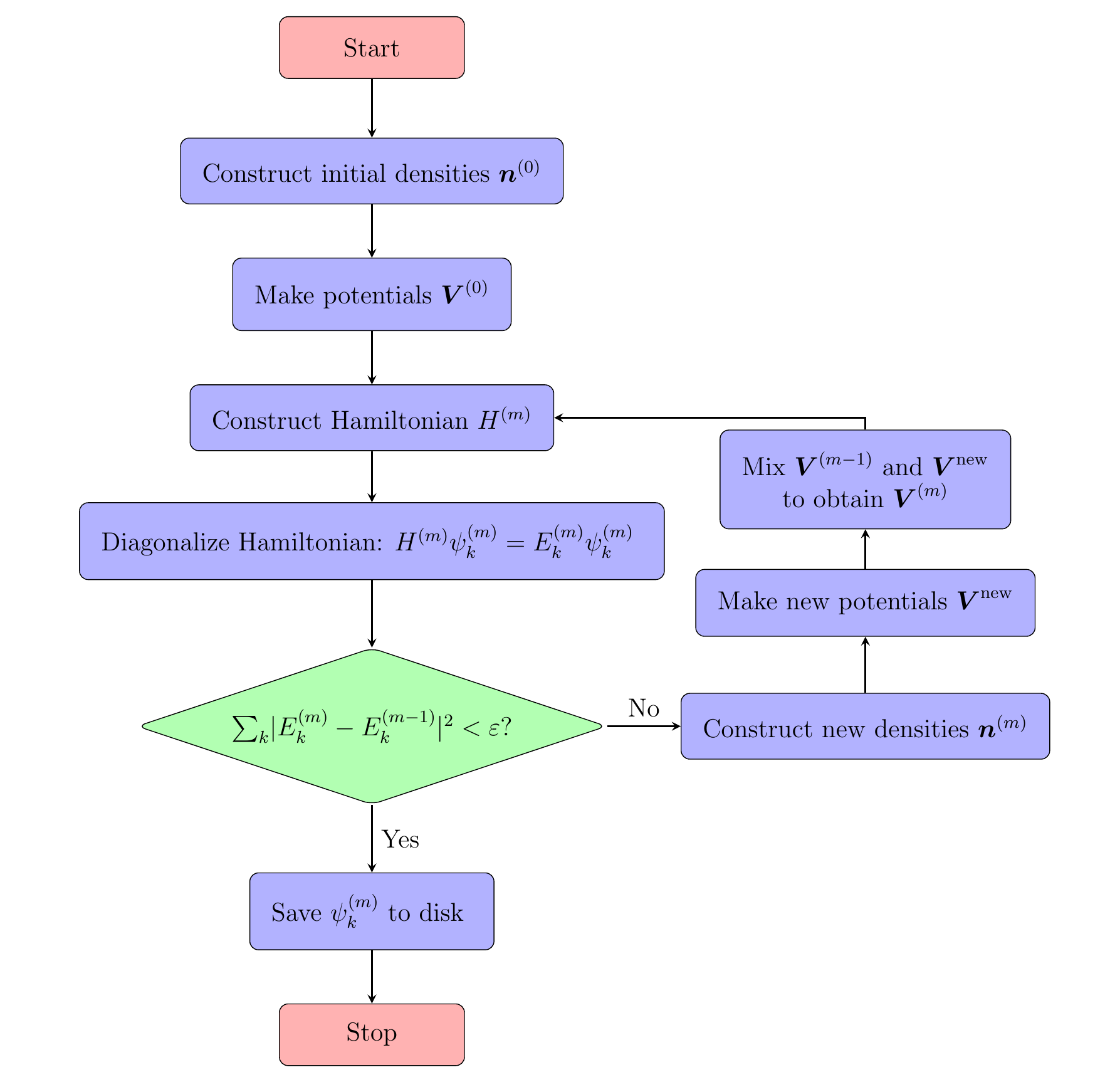}
\caption{ \label{fig:flowchart_slda}
Flowchart of static SLDA solver.
}
\end{figure}

To reach convergence, the new potentials and chemical potentials are mixed with the ones in the last iteration by a linear combination:
\begin{subequations}\label{eq:linear_mixing}
\begin{align}
\vect{V}^{(m)} & = (1-\alpha) \vect{V}^{(m-1)} + \alpha \vect{V}^{\mathrm{new}} \\
\mu^{(m)} & = (1-\alpha) \mu^{(m-1)} + \alpha \mu^{\mathrm{new}}
\end{align}
\end{subequations}
with a constant mixing factor $\alpha$. As an additional option, another mixing procedure is available; it is based on the Broyden method, see Ref.~\cite{Baran:2008} for details. However, we advise caution in using this option, as it often deviates from finding the solution. Note that we concurrently perform interations for proton and neutron density, by splitting the MPI world into two equal groups.

The convergence criteria of the self-consistent iterations is that the norm of the quasiparticle energy difference is less than the tolerance $\varepsilon$:
\begin{align*}
\sum_k \lvert E^{(m)}_{k} - E^{(m-1)}_{k} \rvert^2 < \varepsilon.
\end{align*}
When a converged solution is obtained, the qpwfs are written into files to be used by the time-dependent code. 

\subsubsection{Constraints}

In practical calculations, one often requires a HFB minimum in certain configurations, for example,  multiple mass moments defined in \cref{eq:deform} to have fixed values. Then in a constrained HFB calculation, one needs to add a constraint term into the s.p. Hamiltonian in \cref{eq:spham}
\begin{align}\label{eq:constrained_h}
h' = h + \sum_i  \lambda_i (\hat{Q}_i - Q_i^{(0)}),
\end{align}
where $\hat{Q}_i$'s are different constraint operators, $\lambda_i$s are their corresponding Lagrange multipilers, and $ Q_i^{(0)}$ are the expectation values of operators $\hat{Q}_i$'s input by the user. Between iterations, for a given constraint $\hat{Q}$, the Lagrange multiplier $\lambda$ is updated in the \textit{augmented Lagrangian method} (ALM) \cite{Staszczak:2010}
\begin{align}\label{eq:alm}
\lambda_i^{k+1} = \lambda_i^k + 2c_i(q_i-Q_i^{(0)}),
\end{align}
where $q_i$ is the expectation value of $\hat{Q}_i$ in the iteration $k$. The coefficients $c_i$ should be small enough to guarantee the stability of the self-consistent iterations. In the static code, we implemented the constraints of the center of mass positions of the nucleus $\vec{r}_{\mathrm{cm}}$
and quadruple mass moment $Q_{20}$ with additive operators $\{ \hat{x}, \hat{y}, \hat{z}, \hat{Q}_{20}\}$ associated with the Lagrange multipliers $\{\lambda_x, \lambda_y, \lambda_z, \lambda_{q2} \}$.

\subsection{TDSLDA} \label{sec:algorithm_tdslda} 
\subsubsection{Time evolution}
 A common approach to solving the time-dependent mean-field equations like TDHF is the series expansion method~\cite{Maruhn:2014}. In this method, 
the PDE \ref{eq:tdslda} can be formally rewritten into an integral equation as 
\begin{subequations}
\begin{align}
\psi_k (t+\Delta t)  = \hat{U}(t,t+\Delta t) \psi_k (t),
\end{align}
where $\psi_k(t)$ denotes the qpwf 
$[{\text u}_{k\uparrow}(t),
{\text u}_{k\downarrow}(t),
{\text v}_{k\uparrow}(t),
{\text v}_{k\downarrow}(t)]^T$. The evolution operator $\hat{U}$ is defined as 
\begin{align}
\hat{U}(t,t+\Delta t) = \mathcal{T} \exp\left( -\frac{i}{\hbar} \int_t^{t+\Delta t} \hat{H}(t') dt' \right).
\end{align}
\end{subequations}
where $\mathcal{T}$ the time-ordering operation and $\hat{H}(t)$ is the Hamiltonian at time $t$. For higher accuracy, we have implemented a predictor-corrector method as follows:
\begin{enumerate}
\item At time $t$, we perform a predictor step with the Hamiltonian $H(t)$ constructed from the densities $\vect{n}(t)$ computed using the qpwf $\psi_k(t)$  
\begin{align}
\psi_{\mathrm{pre}} = \exp \left( -\frac{i}{\hbar} \hat{H}(t) \Delta t \right) \psi_k(t).
\end{align}
\item  From the trial solution $\psi_{\mathrm{pred}}$ we obtain a set of predictor densities denoted as $\vect{n}_{\mathrm{pred}}$, that are used to compute a set of corrector densites $\vect{n}_{\mathrm{cor}} = (\vect{n}_{\mathrm{pre}} + \vect{n}(t))/2$. The corrector Hamiltonian $ \hat{H}_{\mathrm{cor}}$ is then constructed with densities $\vect{n}_{\mathrm{cor}}$, and if $\Delta  t$ is small enough,  $ \hat{H}_{\mathrm{cor}}$ is a good approximation to the self-consistent Hamiltonian at time $t + \Delta t /2$.
\item Finally, the qpwf at time $t + \Delta t$ is calculated as
\begin{align}
\psi_k(t+\Delta t) = \exp \left( -\frac{i}{\hbar} \hat{H}_{\mathrm{cor}} \Delta t \right) \psi_k(t).
\end{align}
\end{enumerate}
In our numerical implementation, the  time-evolution operator $\exp \left( -\frac{i}{\hbar} \hat{H} \Delta t \right)$ is replaced with the series expansion
\begin{align}
\exp \left( -\frac{i}{\hbar} \hat{H} \Delta t \right) = \sum_n \frac{(-i)^n}{\hbar^n n!} \hat{H}^n \Delta t^n,
\end{align}
and in practice we find an expansion to order $n=4$ is enough for a good accuracy, with the error of order $\mathcal{O}(\Delta t^5)$. Within an expansion approach, we need to compute $4 \times 2 = 8$ times matrix-vector (MV) products $H\psi$ in each time step, which is the most time-consuming part of the code. 

The Adams-Bashforth-Milne {ABM} method \cite{Hamming:1996} provides an alternative to the 
series expansion method with an increased same accuracy, but only 2 MV products in each time 
step. For a PDE that can be generically written as $y' = f(y)$, in the 5th order predictor-modifier-corrector {ABM} method, the $y_{n+1}$ solution is constructed from 4 previous values as follows:

\pagebreak
\begin{widetext}
\begin{subequations}
\begin{align} 
p_{n+1} & = \frac{y_n+y_{n-1}}{2} + \frac{\Delta t}{48}\left ( {\bf y'_n} - 99 y'_{n-1}+ 69 y'_{n-2} - 17y'_{n-3} \right) + \frac{161}{480}(\Delta t)^5y^{(5)} \label{eq:p}  \\
m_{n+1} & = p_{n+1} - \frac{161}{170}(p_n - c_n) + \frac{923}{2880}(\Delta t)^6y^{(6)} \label{eq:m}\\
c_{n+1} & = \frac{y_n+y_{n-1}}{2} + \frac{\Delta t}{48} \left( 17 {\bf m'_{n+1}} + 51 y'_{n} +3y'_{n-1} + y'_{n-2} \right) - \frac{9}{480} (\Delta t)^5y^{(5)} \label{eq:c}\\
y_{n+1} & = c_{n+1} + \frac{9}{170}(p_{n+1} - c_{n+1}) - \frac{43}{2880}(\Delta t)^6y^{(6)} \label{eq:y}
\end{align}
\end{subequations} \end{widetext}

\noindent
where $p$, $m$, $c$ denote the predictor, modifier and corrector, and a prime 
marks the derivative with respect to time. We specified with bold symbols the two places 
where the quasiparticle Hamiltonian ${\cal H}(\vec{r},t)$, see \cref{eq:tdslda}, 
is applied to the qpwfs per each time-step.
The drawback of the {ABM} method is that it 
cannot self start from step 0 because the time derivatives $y'_n$ at time steps $n=-1, -2, -3$ 
and $y_{-1}$ are unknown, unless the starting state is a stationary solution, and in that case $y'_{-3}=y'_{-2}=y'_{-1}=0$ and $y_{-1}=y_0$. 
However, the series expansion method can serve as a starting procedure for up to four 
steps during which time the derivatives of the wavefunctions are calculated and stored 
in each step to be subsequently used for the ABM method.
 
 
When computing the time evolution of the qpwfs via \cref{eq:tdslda}, an irrelevant phase factor 
 \begin{align} \label{eq:phase}
 \exp \left [ -\frac{i}{\hbar} \int_0^t \langle \psi_k(t') | {\cal H} (t') | \psi_k (t') \rangle dt' \right]
 \end{align}
is introduced.
This factor induces oscillations in time and downgrades the numerical accuracy and stability. 
In our implementation this trivial phase is removed removing the instantaneous quasiparticle energy, 
 \begin{align}
 i\hbar \dot{\psi}_k(t) = [{\cal H} (t)- \eta_k(t))] \psi_k(t) \label{eq:sp_en}
 \end{align}
 with 
 \begin{align}
 \eta_k(t) = \langle \psi_k(t) | {\cal H}(t) | \psi_k (t) \rangle.
 \end{align}
 Here $\psi_k(t)$ stands for the four component qpwf 
 $\left [\text{u}_{k\uparrow} (\vec{r},t), \text{u}_{k\downarrow}(\vec{r},t),  \text{v}_{k\uparrow}(\vec{r},t),\text{v}_{k\downarrow}(\vec{r},t)\right ]^T$.
At the end of each time step we sometimes re-normalize each qpwf $\psi_k$, which results in a negligible overhead, 
but it insures that the numerical roundoff errors are kept to a minimum. We have found however that this renormalization of the qpwfs
is typically not needed.

\subsection{Various numerical tests of the time-dependent code}

\subsubsection{The choice of the spatial lattice constant and of the time-step integration}

After introducing the spatial discretization the emerging time-dependent time-dependent mean field equations constitute a system of 
nonlinear coupled complex partial differential equations (PDEs) with $2\times2\times  N_xN_yN_z\times 4N_xN_yN_z$.  
The first factor 2 stands for the proton and neutron systems. The next factor $2N_xN_yN_z$ stands for the number of quasiparticle states
of the $4N_xN_yN_z\times N_xN_yN_z$ Hartree-Fock-Bogoliubov Hamiltonian. And the last factor $4N_xN_yN_z$ is the number of 
spin and space coordinates in a single quasiparticle wave function. It is trivial to show  and known for decades that by choosing  
the real and imaginary parts of the wave functions at each coordinate the time-dependent mean field equations are formally equivalent
to a non-linear classical Hamiltonian system with  $2\times  N_xN_yN_z\times 4N_xN_yN_z$ degrees of freedom. The only approximation in solving this
large system of non-linear Hamiltonian system, once the size of the simulation box and the spatial lattice constant have been chosen, is the discretization of time. 

Upon discretization one should make sure that various symmetries of the emerging equations of motion are non violated. On a simulation box with 
periodic boundary conditions translational symmetry is trivially satisfied. The isospin symmetry is not violated either after discretization. Gauge invariance, 
corresponding $\psi(\vec{r},t) \rightarrow \exp(if(\vec{r},t)\psi(\vec{r},t)$ is satisfied as well, if one computed various derivatives using the Liebniz rule. 
The local Galilean invariance and the gauge invariance, specific to Bogoliubov mean field, when one applies the operator $\exp(i\hat{N}\eta)$ 
with $\hat{N}$ the number operator,  on the generalized  Slater determinant, are also satisfied   
in the present formulation of the nuclear energy density functional. The only remaining symmetry which is patently broken is the rotational symmetry however. 
We will discuss the time-reversal invariance in the next section.

\begin{figure}[htbp]
\includegraphics[scale=0.4]{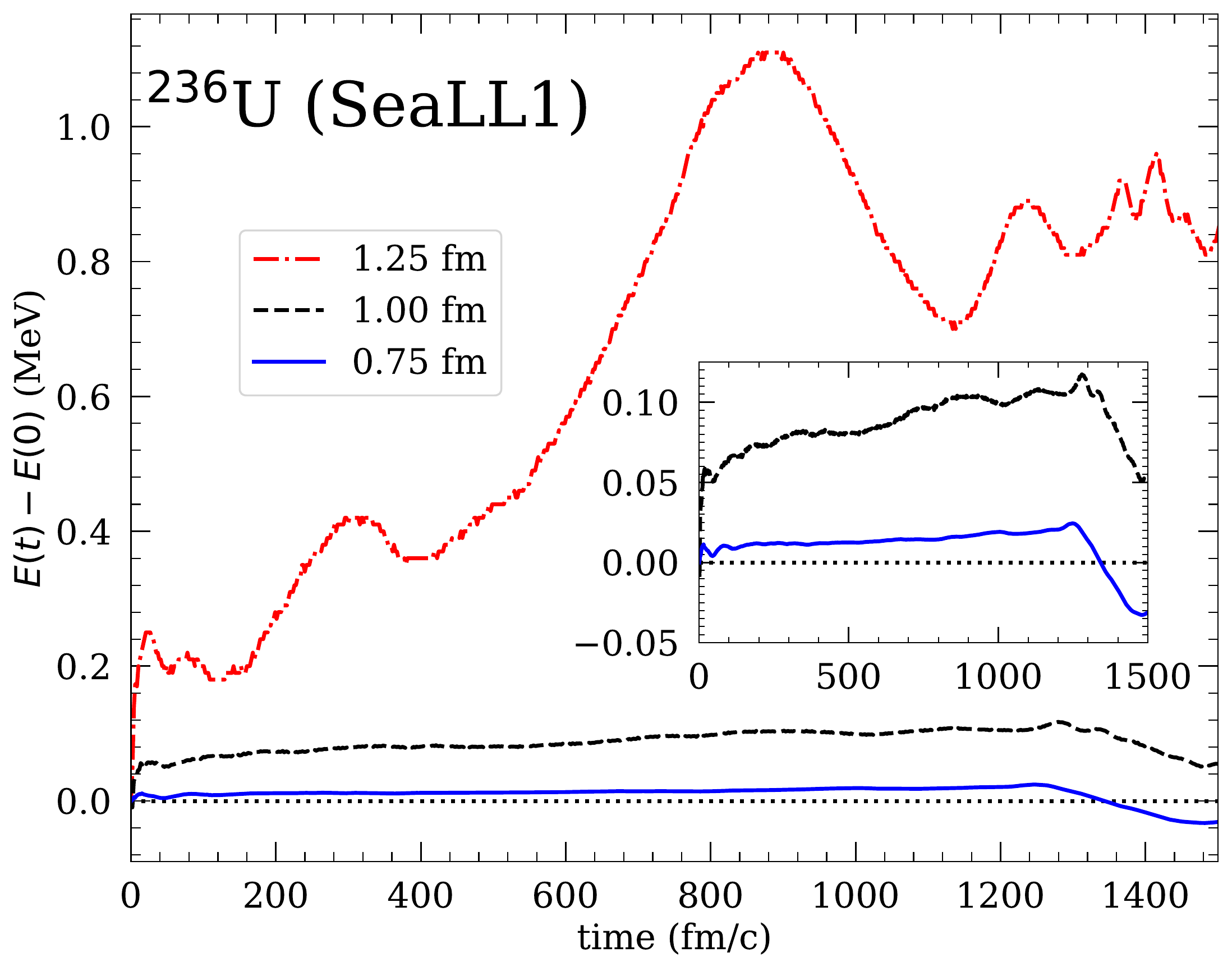}
\caption{ \label{fig:236U}
The evolution with time of the total energy difference $E(t)-E(0)$, which should be conserved exactly if the spatial and time discretization
are adequate.
}
\end{figure}

In almost all published so far simulations of nuclear processes~\cite{Stetcu:2011,Stetcu:2015,Bulgac:2016,Bulgac:2017,Bulgac:2019b,Bulgac:2019c,Bulgac:2020}
we used a spatial lattice constant $a=1.25$ fm, which corresponds to a 1D 
linear momentum cutoff $p_\text{cut}=\hbar\pi/a\approx 496$ MeV/c~\cite{Bulgac:2013x}, comparable in magnitude 
with some of the largest cutoff momenta considered in modern $\chi$-Effective Field Theory models of nucleon-nucleon interactions~\cite{Sammarruca:2019}, 
which are used to describe nucleons interactions in the medium, e.g. the equation of state of pure neutron matter and symmetric nuclear matter. 
Using such a value of the spatial lattice constant leads to a nuclear binding energy of $^{208}$Pb  within the mean field approximation with an accuracy of less 1 MeV~\cite{Ryssens:2015},
if all spatial derivatives are computed using FFT or equivalently Lagrange-mesh techniques. This applies as well to time-dependent simulations, see Fig.~\ref{fig:236U}.

We have performed fission simulations of $^{236}$U with the nuclear energy density functional SeaLL1~\cite{Bulgac:2018a} for various values of the 
spatial lattice constant $a=0.75, 1.00, 1.25$ fm in a simulation box of size $30^3\times60$ fm$^3$. The 1D momentum cutoff for the smaller lattice constants
 $a=0.75, 1.00$ are $p_\text{cut}\approx 827, 620$ MeV/c.   Since we use in 3D a cubic momentum cutoff, see Eq.~\eqref{eq:geff_3D}, the actual value
 of the momentum cutoff in 3D is larger by a factor  of $\sqrt{3}$ than the 1D $p_\text{cut}$ values mentioned above.   
These choices corresponded to evolving in time
$16N_xN_yN_z$ PDEs,  i.e. 2,048,000, 864,000, and 442,368 PDEs respectively, 
on spatial lattices $40^2\times80$, $30^2\times 60$, and $24^2\times 48$ respectively. We observed that the total energy 
is conserved during evolution reasonably well in all these cases. By varying the time-step integration down by a factor of 10 in all cases we 
obtained an identical behavior for each spatial lattice constant. With further tests, by changing the spin-orbit interaction in particular,  
we have identified that the treatment of the spin-orbit interaction alone
is at the root of time time variation of $E(t)-E(0)$, see Fig.~\ref{fig:236U}, along with other changes. 
In particular of the noticeably different spatial-temporal evolutions of the pairing field and 
of the center of mass position of the system in the case of $a=1.25$ fm when compared to $a=1.00, 0.75$ fm.
With decreasing the spatial lattice constant the coarseness of the spatial pixelation is significantly ameliorated and a spatial lattice constant $a=1$ fm 
or less is a satisfactory choice. However, most of other global properties, the total kinetic energy of the fission fragments, their masses and charges, 
and their excitation energies  are little affected by changing the spatial lattice constant from $a=1.25$ to $a=1.00$ fm or less.
The spin-orbit interaction is the only element of the single-particle Hamiltonian which is highly susceptible to rotations, and the only element of the 
single-particle Hamiltonian which performs rotations of the wave functions during evolution, see Eq.~\eqref{eq:spham}. 
The other elements of the single-particle Hamiltonian lead only to 
scale transformations, translations, spin rotations, and boosts along the cartesian axes.  

The time evolution described with the discretization Eqs.~(\ref{eq:p},\ref{eq:m},\ref{eq:c},\ref{eq:y}) is clearly not unitary, but for all the reasonable choices of the time integration 
step the particle number was conserved with high accuracy, $\Delta N(t)/N(0)\ll 10^{-9}$ or even better, even though we typically do not enforce
the normalization of the qpwfs. The integration time step can be determined by educated guess. We determined that $\Delta = 0.06$ fm/c is 
sufficient in the case of $a=1$ fm. Larger time steps can bead to numerical instabilities and smaller time-steps do not lead to any noticeable
numerical improvements. This was a simulation performed in a $30^2\times120$ fm$^3$ simulation box with a spatial lattice constant of $a= 1.25$ fm. 
These types of simulations are important for extracting various moments of observables in the final state, after 
implementing the  Balian and V\'en\'eroni prescription~\cite{Balian:1984}. 

\subsubsection{Time reversal invariance}

\begin{figure}[htbp]
\includegraphics[scale=0.275]{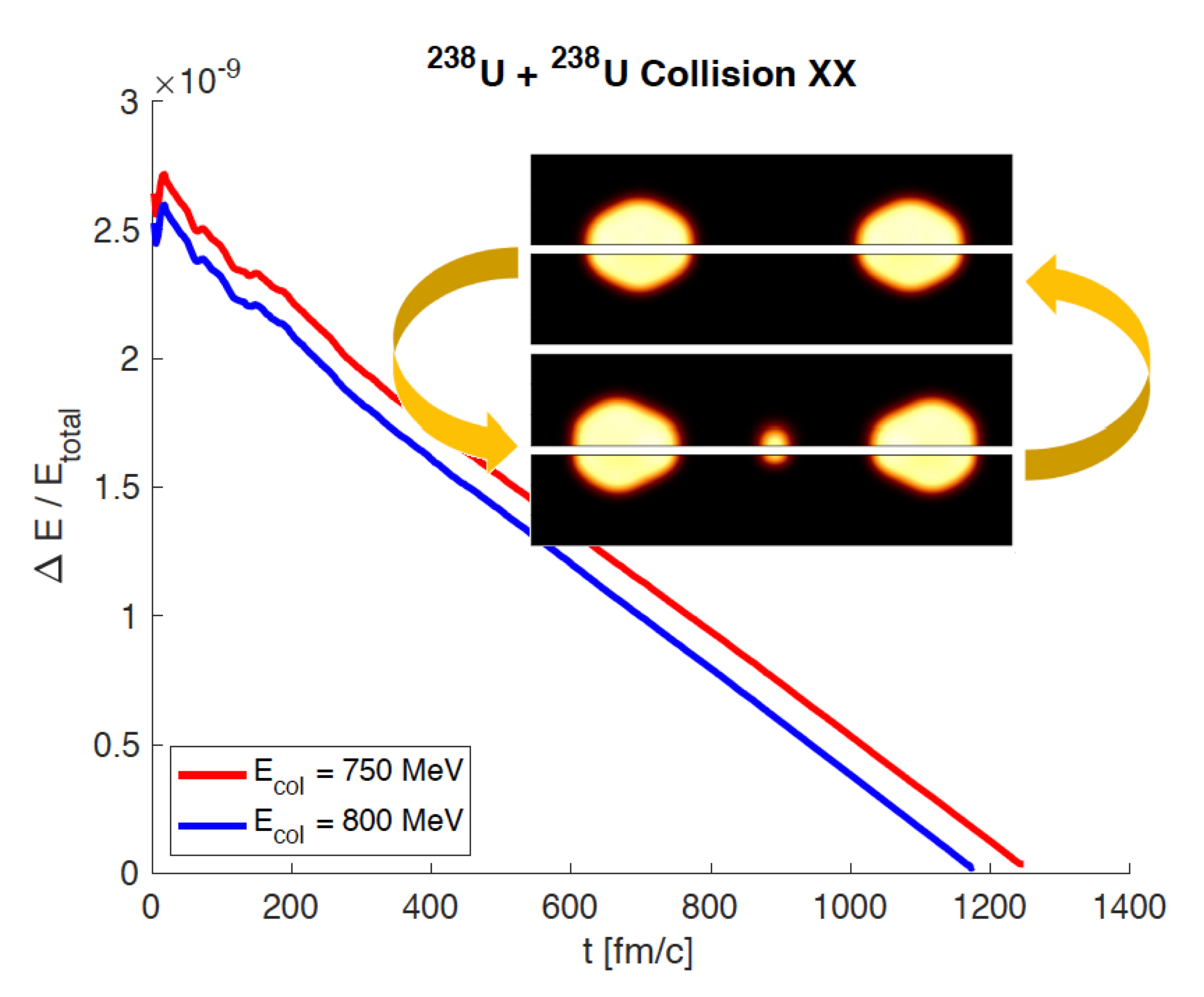}
\caption{ \label{fig:TR}
The head-on collision $^{238}$U+$^{238}$ in the center of mass 
has been performed with the energy density functional SeaLL1 forward and then backward in time.
the inset shows the initial and final neutron and proton densities (in the upper and lower halves of the 
panels of the upper and lower inset). After separating the two fragments leave
a small cluster at rest in middle. At these center of mass collision energies 
a significant part is converted into the internal excitation energies of the fragments.  
The excitation energies of the fragments are in the range $150\ldots 175$ MeV in the case of 800 MeV collision energy. 
}
\end{figure}

It is not obvious that time reversal symmetry would be preserved in the discretized version of TDDFT. In a head-on collision 
of two heavy nuclei, see Fig.~\ref{fig:TR}, we evolved in time the two nuclei first forward in time and subsequently backward
in time towards the initial state. We establish that the total energy of the entire system is recovered with a relative accuracy 
of better than $\approx 3\times 10^{-9}$.  
  
\subsubsection{Numerical ``noise''}

\begin{table*}
\begin{tabular}{ ccccccccccccc}
$\varepsilon$  &$Z_{ini}$& $A_{ini}$ & $E_{ini}$ &$Z_H$&$A_H$&$Z_L$& $A_L$&$Q_{20,H}$  & $Q_{30,H}$ & $Q_{20,L}$ &$Q_{30,L}$&  TKE \\ 
\hline 
                    0& 92.0       & 236.0       &  -1782.9 & 52.2  & 135.7 & 39.8    &100.3 & 2.16            & -0.36         & 14.8            & 0.03          & 171.2\\
       $10^{-8}$& 92.0       & 236.0       &  -1782.9 & 52.2  & 135.7 & 39.8   &100.3 & 2.16            & -0.36          &14.8             & 0.03          & 171.2\\
       $10^{-4}$& 92.0       & 236.0       &  -1782.9 & 52.2  & 135.7 & 39.8   &100.3 & 2.16            & -0.36          & 14.8            & 0.03          & 171.2\\
       $10^{-2}$& 92.0       & 236.0       &  -1780.1 & 52.2  & 135.8 & 39.8   &100.2 & 2.33            & -0.35          & 14.7            & 0.05          & 171.9\\
   $10^{-2}_n$& 92.0       & 236.0       &  -1778.0 & 52.2  & 135.8 & 39.8   &100.2 & 2.33            & -0.35          & 14.7            & 0.05          & 171.9\\
   $3\times10^{-2}$&92.0   &236.1       &  -1758.3 & 52.3  & 135.9  &39.8   & 100.2 & 2.00            &-0.33           & 14.0            & 0.14          & 173.4 \\
\end{tabular}
\caption{\label{tab:noise1} The parameter $\varepsilon$ is the amplitude of the complex ``noise''  defined in Eq.~\eqref{eq:noise}, 
$Z_{ini}, A_{ini}, E_{ini}$ are the initial values of the charge, mass number, and energy of the fissioning $^{236}$U nucleus. 
$Z{H,L}, A_{H,L}, Q_{20,H,L}, Q_{30,H,L}$ are the fission fragments charge, mass number, quadrupole and octupole moments respectively, 
and TKE is the total kinetic energy of the fragments. The multipoles moments are defined as $Q_{20} = 3z^2-r^2$ and $Q_{30}=5z^3-r^2z$ and
measured in barns (b). The energies $E_{ini}$ and TKE are measured in MeVs.
In the simulation corresponding to the 
entrance with $\varepsilon = 10^{-2}_n$ we have normalized the initial qpwfs, but did not orthogonalized them. One can see that a ``noise'' level of $\varepsilon=10^{-2}$ 
the nucleus is excited by approximately $E_{ex}\approx 3$ MeV, thus one has roughly $ \varepsilon ={\cal O}( E_{ex}/|E_{ini}) $ as one might have expected.   }
\end{table*}

The number of real variables we evolve in time in fission studies  ranges from $(4\times N_xN_yN_z)^2\approx 1.22\times10^{10}\ldots 2.62\times 10^{11}$
for the spatial lattices $24^2\times48\ldots 40^2\times80$, and larger in case of collisions of heavy nuclei,  and the number of times steps is of the 
order of ${\cal O}(10^5)\ldots{\cal O}(10^6)$. Even though we perform the calculations with double precision, since the system equations is equivalent to 
a classical non-linear Hamiltonian system with  $(4\times N_xN_yN_z)^2$ coordinates and momenta, and unlike the linear Schr\"odinger equation, such 
a system of equations can be characterized by positive Lyapunov exponents, and thus be chaotic. During the long time evolution either numerical noise 
and roundoff errors can lead potentially to completely unphysical results. 

We performed a number of tests of the following character. We replace the initial qpwfs $\psi_k({\vec r},0)$, see Eqs.~\eqref{eq:tdslda}, at each point in space as follows
\begin{equation}
\tilde{\psi}_k({\vec r},0)=\left(\begin{array}{c}
\text{u}_{k\uparrow}({\vec r},0)[1+\varepsilon\alpha_{k\uparrow}({\vec r})]\\
\text{u}_{k\downarrow}({\vec r},0)[1+\varepsilon\alpha_{k\uparrow}({\vec r})]\\
\text{v}_{k\uparrow}({\vec r},0)[1+\varepsilon\beta_{k\uparrow}({\vec r})]\\
\text{v}_{k\downarrow}({\vec r},0)[1+\varepsilon\beta_{k\uparrow}({\vec r})]
\end{array}\right), \label{eq:noise}
\end{equation}
where we considered $\varepsilon =10^{-8},10^{-4}, 10^{-2}, 10^{-1}$ and 
$\alpha_{k\sigma}({\vec r})$ and $\beta_{k\sigma}({\vec r})$ were chosen as independent complex numbers with the real and 
imaginary parts as random uniform numbers in the interval $(-1,1)$. 
The modified qpwfs $\tilde{\psi}_k({\vec r},0)$ do not form a set on orthogonal vectors in the Fock space. 
We have evolved these initial qpwfs with Eqs.~\eqref{eq:tdslda} and formed the time dependent densities and computed all observables in accord to equations in Section~\ref{sec:theo}.
Even though these type of 
qpwfs are highly unusual, the equations of motion conserve all expected integrals of motion, e.g. total particle number, total energy, total momentum, etc. 
Moreover, this type of ``noise''
strictly speaking destroys the periodicity of our basis functions, see Section~\ref{sec:meshgrid}, and leads to numerical errors beyond those discusses above.
One can in principle mitigate these issues by introducing a filter, which eliminates the 1D momenta $p > p_\text{cut}$. 

This type of stochasticity is similar, though not identical, to the stochastic mean field model  of Ayik~\cite{Ayik:2008,tanimura2017}. 
In the first order in $\varepsilon$, after performing a statistical average, the qpwfs $\tilde{\psi}_k(\vec{r},0)$ are orthogonal to each other and 
the various other observables (particle number, energy, various momenta of the spatial and linear momenta) have values  identical to the unperturbed values, 
but in second order in $\varepsilon$ all statistical averages of observables have non-vanishing values $\propto \epsilon^2$. In particular, even though in all 
our fission studies so far we considered only axially symmetric  initial shapes and vanishing momenta, after introducing noise the all possible 
spatial momenta and momentum momenta are non-vanishing, including varying spatial deformations and non-vanishing velocities for every component of the qpwfs.
We have evolved these initial qpwfs with Eqs.~\eqref{eq:tdslda} and formed the time dependent densities and computed all observables in accord to equations in Section~\ref{sec:theo}.

The results shown in Table~\ref{tab:noise1} are to some extent surprising, as they appear to point to vanishing Lyapunov exponents in TDSLDA simulations. 
As we mentioned above, the TDSLDA equations are equivalent to a classical highly nonlinear mechanical system with an extremely large 
number of coordinates and momenta, in this case $(4\times N_xN_yN_z)^2\approx 4.67\times 10^{10}$. If the Lyapunov exponents would be positive, 
one would expect that the outcome of trajectories started very closely would diverge exponentially. What we see however, that the final properties of 
the fission fragments are only weakly affected. While such an outcome would be expected for a many-body Sch\"odinger equation, which is linear, and 
thus has vanishing Lyapunov exponents, this is unexpected for the time-dependent mean field equations, which are by nature highly non-linear. 
This property of the time-dependent mean field equations is extremely important for applications, if our conclusion is ultimately fully confirmed.


\section{Parallelization and GPU acceleration}
\subsection{Static Code}

\textit{Block-Cylic Distribution} \cite{block_cyclic}.
It matters to understand the storage of the matrix coefficients in computer memory. In general, let $m,n\in\mathbb{Z}^{+}$, then matrix $A\in \mathbb{C}^{m}\otimes \mathbb{C}^{n} \rightarrow  \mathbb{C}^{m\times n}$ is a set of $mn$ complex numbers $\{a_{i,j}\}$ with labels $ i\in [0,m-1]~,~ j\in [0,n-1]$. There are many ways to arrange these coefficients in contiguous memory. We use the column major index map that takes $\{a_{i,j}\}$ to $\{a_{k}\}$ as $(i,j)\rightarrow k=i + j*m$ for all combinations of $i\in[0,m-1], j\in [0,n-1]$ assigning order in memory based on the value of $k$. To recover $(i,j)$ from $k$ use $(\forall k\in [0,mn-1]) k\rightarrow (i=k \mod m, j =\frac{k- (k \mod m)}{m})$. A simple example for the column major layout in memory for $m=9,n=11$ reads:
\begin{align*}
A &=
\begin{pmatrix}
\begin{smallmatrix}
a_{0,0} & a_{0,1} & a_{0,2} & a_{0,3} & a_{0,4} & a_{0,5} & a_{0,6} & a_{0,7} & a_{0,8} & a_{0,9} & a_{0,10}    \\ 
a_{1,0} & a_{1,1} & a_{1,2} & a_{1,3} & a_{1,4} & a_{1,5} & a_{1,6} & a_{1,7} & a_{1,8} & a_{1,9} & a_{1,10}   \\ 
a_{2,0} & a_{2,1} & a_{2,2} & a_{2,3} & a_{2,4} & a_{2,5} & a_{2,6} & a_{2,7} & a_{2,8} & a_{2,9} & a_{2,10}   \\ 
a_{3,0} & a_{3,1} & a_{3,2} & a_{3,3} & a_{3,4} & a_{3,5} & a_{3,6} & a_{3,7} & a_{3,8} & a_{3,9} & a_{3,10}   \\ 
a_{4,0} & a_{4,1} & a_{4,2} & a_{4,3} & a_{4,4} & a_{4,5} & a_{4,6} & a_{4,7} & a_{4,8} & a_{4,9} & a_{4,10}   \\ 
a_{5,0} & a_{5,1} & a_{5,2} & a_{5,3} & a_{5,4} & a_{5,5} & a_{5,6} & a_{5,7} & a_{5,8} & a_{5,9} & a_{5,10}   \\ 
a_{6,0} & a_{6,1} & a_{6,2} & a_{6,3} & a_{6,4} & a_{6,5} & a_{6,6} & a_{6,7} & a_{6,8} & a_{6,9} & a_{6,10}   \\ 
a_{7,0} & a_{7,1} & a_{7,2} & a_{7,3} & a_{7,4} & a_{7,5} & a_{7,6} & a_{7,7} & a_{7,8} & a_{7,9} & a_{7,10}   \\ 
a_{8,0} & a_{8,1} & a_{8,2} & a_{8,3} & a_{8,4} & a_{8,5} & a_{8,6} & a_{8,7} & a_{8,8} & a_{8,9} & a_{8,10}   \\ 
\end{smallmatrix}                     
\end{pmatrix} \\
&=
\begin{pmatrix}
\begin{smallmatrix}
a_{0} & a_{9}   & a_{18} & a_{27} & a_{36} & a_{45} & a_{54} & a_{63} & a_{72} & a_{81} & a_{90} \\
a_{1} & a_{10} & a_{19} & a_{28} & a_{37} & a_{46} & a_{55} & a_{64} & a_{73} & a_{82} & a_{91}   \\ 
a_{2} & a_{11} & a_{20} & a_{29} & a_{38} & a_{47} & a_{56} & a_{65} & a_{74} & a_{83} & a_{92}   \\ 
a_{3} & a_{12} & a_{21} & a_{30} & a_{39} & a_{48} & a_{57} & a_{66} & a_{75} & a_{84} & a_{93}  \\ 
a_{4} & a_{13} & a_{22} & a_{31} & a_{40} & a_{49} & a_{58} & a_{67} & a_{76} & a_{85} & a_{94}   \\ 
a_{5} & a_{14} & a_{23} & a_{32} & a_{41} & a_{50} & a_{59} & a_{68} & a_{77} & a_{86} & a_{95}  \\ 
a_{6} & a_{15} & a_{24} & a_{33} & a_{42} & a_{51} & a_{60} & a_{69} & a_{78} & a_{87} & a_{96}   \\ 
a_{7} & a_{16} & a_{25} & a_{34} & a_{43} & a_{52} & a_{61} & a_{70} & a_{79} & a_{88} & a_{97}   \\ 
a_{8} & a_{17} & a_{26} & a_{35} & a_{44} & a_{53} & a_{62} & a_{71} & a_{80} & a_{89} & a_{98}  \\ 
\end{smallmatrix}  
\end{pmatrix} \\
&=\{a_0, a_1, a_2, \dots, a_{97},a_{98}\}.
\end{align*}

In static SLDA calculations, the Hamiltonian matrix $H\in \mathbb{C}^{N\times N} = \mathbb{C}^{4 N_x N_y N_z\times 4 N_x N_y N_z}$ must be diagonalized for each species. In practical calculations, where the dimension $N$ is $10^5 \sim 10^6$, the storage of the HFB matrix requires $O(2^{40})$ bytes, which can easily exceed the maximum memory of a single CPU. To overcome this memory issue and to target parallel processing we use MPI, the message passing interface library, to organize our problem in distributed memory. We split the global MPI communicator into two distinct communicator spaces, i.e. protons and neutrons, and assign $np$ processes to each. In general, the matrix coefficients of $A\in \mathbb{C}^{m\times n}$ can be assigned to a set of $np$ processes in distributed memory labeled by process ids $iam=0,...,np-1$, using the column major 2d block cyclic mapping. Let $np=p\times q$, then each process $iam$ is assigned to a pair of indices that label the coordinates in a rectangular array of processes $(ip,iq) \ni ip=0,..,p-1$ and $iq=0,...,q-1$, i.e. 
\begin{align*}
ip &= iam \mod p ~,\\
iq &=\frac{iam - (iam \mod p) }{p}. 
\end{align*}
The portion of $A$ assigned to process $(ip,iq)$, $A_{ip,iq}$, has memory requirements for
$m_{ip}\times n_{iq}$ coefficients where 
\begin{align*}
m_{ip} &= (\frac{m}{p\cdot mb})\cdot mb ~,\\ 
 &ip<\frac{m \mod (p\cdot mb)}{mb} \rightarrow m_{ip} \mathrel{+}= mb ~,\\
 &ip=\frac{m \mod (p\cdot mb)}{mb} \rightarrow m_{ip} \mathrel{+}= m \mod mb 
\end{align*}
and 
\begin{align*}
n_{iq} &= (\frac{n}{q\cdot nb})\cdot nb ~,\\
 &iq<\frac{n \mod (q\cdot nb)}{nb} \rightarrow n_{iq} \mathrel{+}= nb ~,\\
 &iq=\frac{n \mod (q\cdot nb)}{nb} \rightarrow n_{iq} \mathrel{+}= n \mod nb. 
\end{align*}
The parameters for the map are the matrix dimensions $m,n$, the block sizes for each dimension $mb,nb$ to cyclically map the indices $(i,j)$ of $A$ to local indices $(i_{ip},j_{iq})$ of $A_{ip,iq}$, the process grid parameters $p,q$ and labels $(ip,iq)$. Given $(i,j)$, process $(ip,iq)=( (\frac{i}{mb})\mod p, (\frac{j}{nb})\mod q)$ is the process assigned the element such that $(i,j)\rightarrow (i_{ip},j_{iq})$ where
\begin{align*}
 i_{ip} &= (\frac{i}{p\cdot mb})\cdot mb + i \mod mb ~,\\ 
 j_{iq} &= (\frac{j}{q\cdot nb})\cdot nb + j \mod nb ~. 
 \end{align*}
Last, the column major map is used to assign the local pair of indices to the single local index $k_{(ip,iq)}=i_{ip} + j_{iq} m_{ip}, k_{(ip,iq)} \in [0, m_{ip}n_{iq}-1]$.
Applying the map for the example $m=9,n=11$ matrix, when $p=2,q=3,mb=3,nb=2$ and subscripts labeling $(ip,iq)$, one finds:
$$
\begin{pmatrix} 
\begin{smallmatrix}
\left(\begin{smallmatrix}
a_{0,0} & a_{0,1}  & a_{0,6} & a_{0,7}  \\  
a_{1,0} & a_{1,1}  & a_{1,6} & a_{1,7} \\
a_{2,0} & a_{2,1}  & a_{2,6} & a_{2,7} \\
a_{6,0} & a_{6,1} & a_{6,6} & a_{6,7} \\
a_{7,0} & a_{7,1} & a_{7,6} & a_{7,7} \\
a_{8,0} & a_{8,1} & a_{8,6} & a_{8,7} \\
\end{smallmatrix}\right)_{0,0}

&
\left(\begin{smallmatrix}
a_{0,2} & a_{0,3} & a_{0,8} & a_{0,9} \\
a_{1,2} & a_{1,3} & a_{1,8} & a_{1,9} \\
a_{2,2} & a_{2,3} & a_{2,8} & a_{2,9} \\
a_{6,2} & a_{6,3} & a_{6,8} & a_{6,9} \\
a_{7,2} & a_{7,3} & a_{7,8} & a_{7,9}  \\
a_{8,2} & a_{8,3} & a_{8,8} & a_{8,9} \\
\end{smallmatrix}\right)_{0,1}

&
\left(\begin{smallmatrix}
a_{0,4} & a_{0,5} & a_{0,10}  \\  
a_{1,4} & a_{1,5} & a_{1,10}  \\
a_{2,4} & a_{2,5} & a_{2,10}   \\
a_{6,4} & a_{6,5} & a_{6,10}  \\
a_{7,4} & a_{7,5} & a_{7,10}   \\ 
a_{8,4} & a_{8,5} & a_{8,10}   \\ 
\end{smallmatrix}\right)_{0,2}
\\
\left(\begin{smallmatrix}
a_{3,0} & a_{3,1} & a_{3,6} & a_{3,7}   \\ 
a_{4,0} & a_{4,1} & a_{4,6} & a_{4,7}    \\ 
a_{5,0} & a_{5,1} & a_{5,6} & a_{5,7}    \\ 
\end{smallmatrix}\right)_{1,0}

&
\left(\begin{smallmatrix}
a_{3,2} & a_{3,3} & a_{3,8} & a_{3,9}   \\ 
a_{4,2} & a_{4,3} & a_{4,8} & a_{4,9}    \\ 
a_{5,2} & a_{5,3} & a_{5,8} & a_{5,9}   \\ 
\end{smallmatrix}\right)_{1,1}

&
\left(\begin{smallmatrix}
a_{3,4} & a_{3,5} & a_{3,10}   \\ 
a_{4,4} & a_{4,5} & a_{4,10}   \\ 
a_{5,4} & a_{5,5} & a_{5,10}   \\ 
\end{smallmatrix}\right)_{1,2}

\end{smallmatrix}
\end{pmatrix} .
$$
The reverse map takes $(i_{ip},j_{iq})\rightarrow (i,j)$ where
\begin{align*}
i &= i_{ip} \mod mb + (\frac{i_{ip}}{mb})\cdot p\cdot mb + ip\cdot mb ~,\\ 
j &= j_{iq} \mod nb + (\frac{j_{iq}}{nb})\cdot q\cdot nb + iq\cdot nb.
\end{align*}
Continuing the example, here the matrix $A_{0,0}$ locally managed by process $(ip,iq)=(0,0)$ is mapped back the column major position in the original matrix $A$:
$$
\left(\begin{smallmatrix}
a_{0,0} & a_{0,1}  & a_{0,6} & a_{0,7}  \\  
a_{1,0} & a_{1,1}  & a_{1,6} & a_{1,7} \\
a_{2,0} & a_{2,1}  & a_{2,6} & a_{2,7} \\
a_{6,0} & a_{6,1} & a_{6,6} & a_{6,7} \\
a_{7,0} & a_{7,1} & a_{7,6} & a_{7,7} \\
a_{8,0} & a_{8,1} & a_{8,6} & a_{8,7} \\
\end{smallmatrix}\right)_{0,0}
\overset{2dbc^{-1}}\rightarrow
\begin{pmatrix}
\begin{smallmatrix}
a_{0,0} & a_{0,1}  & * & * & * & * & a_{0,6} & a_{0,7} & *& * & *    \\ 
a_{1,0} & a_{1,1}  & * & * & * & * & a_{1,6} & a_{1,7} & *& * & *   \\ 
a_{2,0} & a_{2,1}  & *& * & * & * & a_{2,6} & a_{2,7}& * & * & *  \\ 
* & * & * & * & * & * & * & * & * &* & *  \\ 
* & * & * & * & * & * & * & * & * &* & *  \\ 
* & * & * & * & * & * & * & * & * &* & *  \\ 
a_{6,0} & a_{6,1}  & *& * & * & * & a_{6,6} & a_{6,7} & *& * & *   \\ 
a_{7,0} & a_{7,1}  & *& * & * & * & a_{7,6} & a_{7,7} & *& * & *   \\ 
a_{8,0} & a_{8,1}  & *& * & * & * & a_{8,6} & a_{8,7} & *& * & *  \\ 
\end{smallmatrix}                     
\end{pmatrix} .
$$ 

%
%
%
%

The diagonalization of the distributed 2d block-cyclic matrix is performed using the \texttt{pzheevd()} routine in the ScaLAPACK library \cite{scalapack}. After the diagonalization, the resulting eigenvectors (qpwfs) are distributed in different processes as a matrix in the same way as the Hamiltonian. To calculate the various densities, one needs to perform sums over qpwfs as shown in \cref{eq:local_densities}. Looping over the global column indices $j$ of the eigenvector matrix, processes with $iq=(\frac{j}{nb})\mod q$ locally compute and copy the portion of the vector in local memory to a full vector buffer, and use MPI collective operations to construct full vectors in each group. The full vectors are necessary to evaluate derivatives of the qpwfs, which enter in the calculation of different densities using the \texttt{MPI\_AllReduce()} routine. The local proton densities calculated in the proton MPI group are shared with the neutron MPI group, and vice versa.  The density exchange is realized via point-point communications between MPI groups using the \texttt{MPI\_Send()} and \texttt{MPI\_Recv()} routines.



\begin{figure}[htbp]
\includegraphics[scale=0.5]{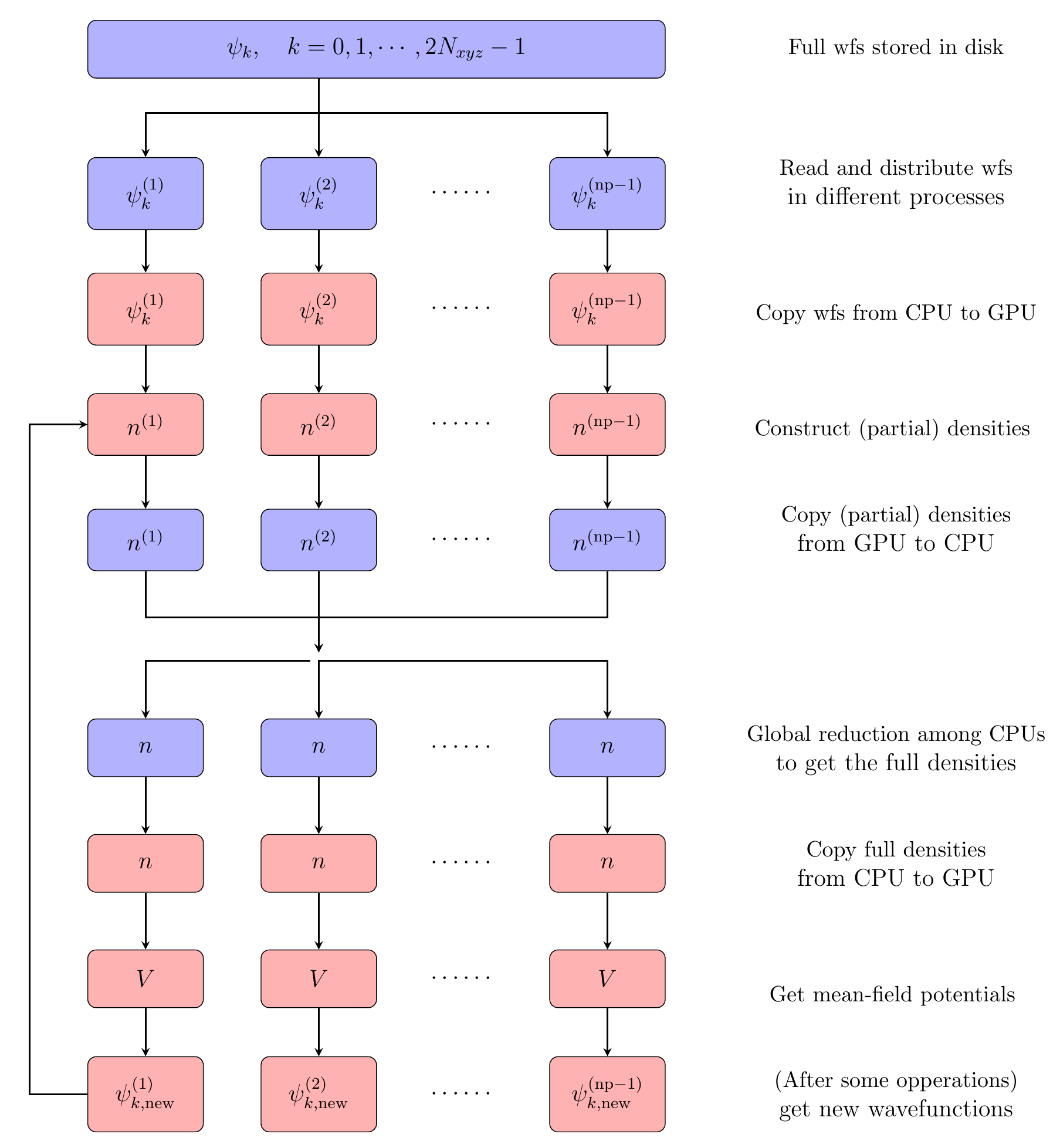}
\caption{
 \label{fig:flowchart_tdslda}
 Parallelization structure of the TDSLDA code.
}
\end{figure}

\subsection{Time-dependent Code}

\subsubsection{CPU-GPU hybrid structure}
The TD code has a straightforward parallelization structure  (illustrated in \cref{fig:flowchart_tdslda}), which is simpler than for the static code and takes advantage of the GPU acceleration.  At the start of the calculation, all qpwfs stored on disk are read and distributed to different (CPU) processes uniformly, i.e. each process receives a certain number \texttt{nwfip} of four-component qpwfs. The size of \texttt{nwfip} depends on the number of qpwfs that has to be evolved, and the number of processes available for the calculation. The qpwfs are copied to the paired GPU device in each process, where most of the computation is performed. Since $nwfip$ qpwfs are available on each GPU, partial densities are constructed first on the GPU, then these partial densities are transferred to the CPU. The total densities are contructed using a global reduction operation \texttt{MPI\_Allreduce}, followed by a copy of the full densities back to each GPU. On each GPU, the mean-field potentials are calculated and then used to evolve the qpwfs in time. Note that since different gradients of the qpwfs are used both in the construction of densities and during the evolution, such quantities are calculated only once and stored in temporary arrays until no longer necessary.

\subsubsection{FFT in batch}
The calculation of the spatial derivatives of qpwfs is the most time-consuming part of the TD code. For such operations, we use  the Nvidia CUDA FFT library for GPUs cuFFT \cite{cufft}, and, in order to take full advantage of the GPU acceleration, we perform FFTs (and inverse FFTs) for a batch of functions simultaneously. In the code, the wavefunctions are stored in a 1-D array of size \texttt{4*nxyz*nwfip*sizeof(cufftDoubleComplex)}. Currently, the value of \texttt{batch} is set to 8, but with the development of the GPU hardware this number, and the number of threads per thread block, can be modified to optimize the performance.
 
\subsubsection{Checkpoint and restart}
During the evolution, for every \texttt{time\_cp} time steps (in fm/c) the wavefunctions are copied from GPUs to CPUs and then saved to disk for a possible restart if evaluation is interrupted for any reason. The same checkpoint procedure is performed at the end of the program as well. During the checkpoint, only the latest wavefunctions (for neutron and proton) are saved and the program can restart from the latest saved time using the series expansion of the time evolution operator described in \cref{sec:algorithm_tdslda}. 


\section{Code Description}

\subsection{Static Code}
The static code contains the following source files:
\begin{itemize}
\item
\texttt{cnuclear-slda-solver\_pn.c}: the main program
\begin{itemize}
\item
\texttt{main()}: the main function;
\item
\texttt{readcmd()}: detects the availability of the input file;
\item
\texttt{parse\_input\_file()}: reads in the input options from a file;
\end{itemize}

\item
\texttt{create\_destroy\_mpi\_groups.c}:
\begin{itemize}
\item
\texttt{create\_mpi\_groups()}: splits the MPI space into two equal spaces, one for protons, one for neutrons;
\item
\texttt{destroy\_mpi\_groups()}: destroys the MPI groups;
\end{itemize}

\item 
\texttt{constr\_dens.c}: contains functions related to densities
\begin{itemize}
\item
\texttt{make\_coordinates}(): constructs the variables associated with the lattice;
\item
\texttt{generate\_ke\_1d}: constructs the DVR of second order derivative in 1D;
\item
\texttt{generate\_der\_1d}: constructs the DVR of first order derivative in 1D;
\item
\texttt{grid3}: returns the corresponding 3D meshgrid from 1D mesh grid;
\item
\texttt{compute\_densities}():  from the eigenvectors of the HFB Hamiltonian (see \texttt{make\_ham}), the densities are constructed at zero temperature;
\item
\texttt{exch\_nucl\_dens()}: exchange of densities and currents between proton and neutron spaces;
\item
\texttt{rescale\_dens()}: option to rescale all the densities to the correct number of particles;
\end{itemize}

\item
\texttt{operators.c}: contains miscellaneous tool functions
\begin{itemize}
\item
\texttt{gradient\_real()}: on the given lattice, compute the gradient of a function via Fourier transforms - the input and output are real;
\item
\texttt{gradient()}: the same as \texttt{gradient\_real()}, but the input and output are complex;
\item
\texttt{gradient\_ud()}: the same as \texttt{gradient()}, but for a 2-component (spin-up and down) function;
\item
\texttt{laplacean()}: computes the laplacean of a real function;
\item
\texttt{match\_lattices()}: computes offsets so that a small lattice is placed in the middle of another lattice with sides three times the largest dimension of the smaller lattice. This is used in calculation of the Coulomb interaction in order to remove influence of image charges;
\end{itemize}

\item
\texttt{make\_potentials.c} contains functions related to the mean-field potentials 
\begin{itemize}
\item
\texttt{dens\_func\_params()}: based on the input option \texttt{force}, sets the parameters of a Skyrme-type interaction (the default is the SLY4 interaction);
\item
\texttt{get\_u\_re()}: computes the real part of the self-consistent potential;
\item
\texttt{coul\_pot3()}: compute the Coulomb potential;
\item
\texttt{update\_potentials()}: computes all the self-consistent potentials from densities and current densities;
\item
\texttt{mix\_potentials()}: mixes the previous potential with the current one, to construct the potentials that will be used in the next iteration;
\item
\texttt{center\_dist}: calculates the center of mass for a given density;
\end{itemize}

\item
\texttt{ham\_matrix.c}: contains functions related to the construction of HFB hamiltonian matrix
\begin{itemize}
\item
\texttt{make\_ham()}: computes the proton/neutron Hamiltonians in each space, cyclically decomposed on a grid of processes. Each Hamiltonian will be diagonalized using the \texttt{pzheevd\_()} routine from ScaLAPACK;
\end{itemize}

\item
\texttt{system\_energy.c}: contains functions related to the calculation of energies
\begin{itemize}
\item
\texttt{system\_energy()}: computes the energy of the nucleus given the computed densities;
\end{itemize}

\item
\texttt{dens\_io.c}: contains functions related to the I/O of densities
\begin{itemize}
\item
\texttt{read\_dens()}: reads from the disk a set of previously computed densities;
\item
\texttt{write\_dens()}: saves the current proton/neutron densities on the disk;
\end{itemize}

\item
\texttt{broyden\_min.c}: contains functions related to the Broyden mixing procedure
\begin{itemize}
\item
\texttt{broydenMod\_min()}: applies the Broyden mixing technique to the current and previous set of potentials;
\end{itemize}

\item
\texttt{dens\_start.c}: contains functions related to generating the initial guess of densities
\begin{itemize}
\item
\texttt{dens\_startTheta()}: constructs guess densities for proton and neutrons;
\end{itemize}

\item
\texttt{deform.c}: contains functions related to the deformation properties of nucleus
\begin{itemize}
\item
\texttt{deform()}: computes the deformation parameters;
\end{itemize}
\item
\texttt{get-blcs-dscr.c, get-mem-req-blk-cyc.c}: contains functions related to the BLACS descriptor of the 2D block-cyclic distribution

\item
\texttt{2dbc-slda-mpi-wr.c}: contains function for writing the qpwfs to a single file on disk from a 2D block-cyclic matrix in distributed memory using MPI IO semantics for portability 
\begin{itemize}
\item
\texttt{bc\_wr\_mpi()}: function that uses MPI IO to write ($q$ write processes, the number of columns in the $p\times q$ process grid) a range of columns from a 2D block-cyclic matrix in distributed memory to a single file on disk (recommended);
\end{itemize}

\item
\texttt{bc-wr-lstr-ec.c}: contains functions for writing the qpwfs to disk in a Lustre file system
\begin{itemize}
\item
\texttt{bc\_wr\_lstr\_ec()}: writes the qpwfs to disks in Lustre file system;
\end{itemize}

\item
\texttt{print\_wf.c}: contains functions for writing the qpwfs to disk with Unix \texttt{write()} 
\begin{itemize}
\item
\texttt{print\_wf()}: writes all qpwfs to disks with Unix \texttt{write()};
\item
\texttt{print\_wf2()}: same as \texttt{print\_wf()}, but only writes qpwfs whose occupation number $>$ 0.9 to disks (usually used for Hartree-Fock calculation, i.e. the pairing is missing).
\end{itemize}
\end{itemize}


\subsection{Time-dependent Code}
\subsubsection{Source files}
The TD code contains the following source files:
\begin{itemize}

\item
\texttt{ctdslda.c}: the main program

\item
\texttt{nuclear-gpu.c} the CUDA source file, contains all the functions related to runnings on GPUs -among them the following routines are the same with the ones in the static code, but in GPU version:
\begin{itemize}
\item
\texttt{compute\_densities\_gpu()};
\item
\texttt{get\_u\_re\_gpu()};
\item
\texttt{update\_potentials\_gpu()};
\item
\texttt{do\_get\_coulomb()};
\end{itemize}
Other new routines are:
\begin{itemize}
\item
\texttt{tstep\_gpu()}: performs the series expansion for 1 time step;
\item
\texttt{get\_hpsi\_gpu()}: performs one time $H\psi$ operation;
\item
\texttt{adams\_bashforth\_pm\_gpu()}: performs the first two lines of ABM method;
\item
\texttt{adams\_bashforth\_cy\_gpu()}: performs the last two lines of ABM method;
\item
\texttt{adams\_bashforth\_dfdt\_gpu()}: calculates the time-derivative of qpwfs in each time step;
\item
\texttt{do\_get\_gradient\_laplacean()}: calculates the gradient and laplacean of qpwfs;
\end{itemize}

\item
\texttt{operators.c}: the same as the one in static code

\item
\texttt{densities.c}: the same as the \texttt{constr\_dens.c} in static code

\item
\texttt{system\_energy.c}: the same as the one in static code

\item
\texttt{deform.c}: the same as the one in static code

\item
\texttt{rotation.c}: contains CPU functions related the rotation properties of the nucleus on the lattice

\item

\texttt{wf.c}: contains functions that handle qpwfs (allocate memory, initialization and i/o). The most important functions included are: 
\begin{itemize}
\item
\texttt{read\_wf\_MPI()}: function that uses MPI IO to read the qpwfs from single file on disk to a set of distributed memory processes (recommended);

\item
\texttt{write\_wf\_MPI()}: function that uses MPI IO to write the qpwfs from distributed memory to a single file on disk -can be used for checkpoint and restart processes (recommended);
\end{itemize}

\item
\texttt{wf-cpt-lstr.c}: similar to \texttt{bc-wr-lstr-ec.c} in static code, contains functions of writing the qpwfs to disks with Lustre library. 
\begin{itemize}
\item
\texttt{wf\_cpt\_lstr()}: writes qpwfs into disk with Lustre library.
\end{itemize}
\item
\texttt{wf-rd-lstr.c}: contains functions of reading the qpwfs from disks with Lustre library. 
\begin{itemize}
\item
\texttt{wf\_rd\_lstr()}: reads the qpwfs from disks with Lustre library.
\end{itemize}

\end{itemize}

\subsubsection{Important variables}
In the TD code all the qpwfs, densities, and potentials are stored as 1D arrays in GPUs. Here we provide a list of important variables on GPU and their storage structure.
\begin{itemize}
\item
\texttt{d\_wavf:} wavefunctions (GPU), length: \texttt{8*Nxyz*nwfip}.

\item
\texttt{d\_wavf\_td:} time derivatives of wavefunctions, length: \texttt{16*Nxyz*nwfip}.

\item
\texttt{d\_wavf\_p:} predictor of wavefunctions, length: \texttt{8*Nxyz*nwfip}.

\item
\texttt{d\_wavf\_c:} corrector of wavefunctions, length: \texttt{8*Nxyz*nwfip}.

\item
\texttt{d\_wavf\_m:} modifier of wavefunctions , length: \texttt{4*Nxyz*nwfip}.

\item
\texttt{d\_densities:} local densities (for each neutron and proton). The arrangement of 
components is listed in \cref{tab: mem_dens}. Total length: \texttt{14*Nxyz}
\begin{table}[htbp]
\center
\begin{tabular}{c|c}
\hline\hline

offset (in double) & density \\
\hline
0               & $\n(\vec{r})$\\   
\texttt{Nxyz}   & $\tau(\vec{r})$\\
\texttt{2*Nxyz} & $\vec{s}(\vec{r})$  \\
\texttt{5*Nxyz} & $ \vec{\nabla} \cdot \vec{J}(\vec{r})$  \\
\texttt{6*Nxyz} & $\vec{j}(\vec{r})$  \\
\texttt{9*Nxyz} & $ \vec{\nabla} \times \vec{j}(\vec{r})$  \\
\texttt{12*Nxyz} & $\kappa(\vec{r})$  \\
\hline\hline
\end{tabular}
\caption{\label{tab: mem_dens}
memory arrangement of \texttt{d\_densities}
}
\end{table}

\item
\texttt{d\_potentials:} local potentials (for each neutron and proton). The arrangement of 
components is listed in \cref{tab: mem_pot}. Total length: \texttt{16*Nxyz+6}
\end{itemize}

\begin{table}[htbp]
\center
\begin{tabular}{c|c|c}
\hline\hline

offset (in double) & potentials & definition\\
\hline
0               & $U(\vec{r})$  & \cref{eq:u}\\   
\texttt{Nxyz}   & $\frac{\hbar^2}{2m^*(\vec{r})}$ & \cref{eq:mass_eff}\\   
\texttt{2*Nxyz}   & $\nabla^2 \frac{\hbar^2}{2m^*(\vec{r})}$ &\\
\texttt{3*Nxyz} & $\vec{W}(\vec{r})$  & \cref{eq:spin_orbit_u}\\
\texttt{6*Nxyz} & $\vec{S}(\vec{r})$  & \cref{eq:Sq} \\
\texttt{9*Nxyz} & $\vec{A}(\vec{r})$  &  \cref{eq:Aq} \\
\texttt{12*Nxyz} & $\Delta(\vec{r})$  & \cref{eq:pairingfield} \\
\texttt{14*Nxyz} & $U_{\mathrm{ext}}(\vec{r})$ & \\
\texttt{15*Nxyz} & $U_{\mathrm{constr}}(\vec{r})$ &\\
\texttt{16*Nxyz} & $\vec{v}_{\mathrm{cm}}$ & \cref{eq:vel}\\
\texttt{16*Nxyz+3} & $\vec{\omega}$ & \\
\hline\hline
\end{tabular}
\caption{\label{tab: mem_pot}
memory arrangement of \texttt{d\_potentials}
}
\end{table}

\section{Input and output description} \label{sec:io}
\subsection{Static code}
\subsubsection{Input}
In the static code, one needs to pass an input file to the executable. An example is the following \texttt{input.test.txt} file
\begin{verbatim}
nx 24
ny 24
nz 48
dx 1.25
dy 1.25
dz 1.25
broyden 0
niter 1
N 146
Z 94
iext 0
force 1
pairing 1
alpha_mix 0.25
ecut 100.0
irun 0
print_wf 0
deform 0
p 32
q 48
mb 40
nb 40
\end{verbatim}
with datatypes and definitions of options:

\begin{itemize}
\item
\texttt{nx, ny, nz}: Integers, the lattice numbers in each direction.

\item
\texttt{dx, dy, dz}: Double, the lattice constants in each direction.

\item
\texttt{N,Z}: Integers, the preset neutron and proton number of nucleus.

\item
\texttt{force}: Integer, the type of force  (NEDF). Available options are:
\begin{itemize}
\item
\texttt{iforce = 1}: SLy4 \cite{Chabanat:1998} EDF with volume pairing.
\item
\texttt{iforce = 11}: SLy4 EDF with mixed pairing.
\item
\texttt{iforce = 12}: SLy4 EDF with surface pairing.
\item
\texttt{iforce = 13}: SLy4 EDF with 0.25 volume plus 0.75 surface pairing.
\item
\texttt{iforce = 2}: SkP \cite{dobaczewski1984} EDF with volume pairing.
\item
\texttt{iforce = 3}: SkM* \cite{Bartel:1982} EDF with mixed pairing, same bare pairing coupling $g_0$ for neutron and proton.
\item
\texttt{iforce = 4}: SkM* EDF with mixed pairing, different bare pairing coupling $g_0$ for neutron and proton.
\item
\texttt{iforce = 5}: SLy5 \cite{Chabanat:1998} EDF with volume pairing.
\item
\texttt{iforce = 6}: SLy6 \cite{Chabanat:1998} EDF with volume pairing.
\item
\texttt{iforce = 7}: SeaLL1 \cite{Bulgac:2018a} EDF with volume pairing.
\end{itemize}

\item
\texttt{p,q}: Integers, the dimension of 2D CPU grid. The product of \texttt{p} and \texttt{q} must equal to \texttt{np/2} where \texttt{np} is the total number of MPI processes.

\item
\texttt{mb,nb}: Integers, the dimension of the block. Suggested values are \texttt{mb=nb=40}

\item
\texttt{ecut}: Double, the energy cutoff in MeV (only applicable for spherical cutoff). Suggested value is 100.

\item
\texttt{pairing}: Integer. Pairing is turned off when \texttt{pairing = 0}.

\item
\texttt{broyden}: Integer. Broyden mixing is used when \texttt{broyden=1}, otherwise linear mixing is used.

\item
\texttt{irun}: Integer. Iterations starts from scratch densities when \texttt{irun = 0}; starts from existing densities when \texttt{irun = 1} .
\item
\texttt{iext}: Integer. The external potential is added when \texttt{iext = 1}.

\item
\texttt{alpha\_mix}: Double, the mixing factor of linear mixing. Suggested value is 0.25.

\item
\texttt{deform}: Integer, applicable for \texttt{irun = 0} case only: start from spherical initial densities when \texttt{deform = 0}; start from quadruple deformed initial densities when \texttt{deform = 1}; start from tri-axial initial densities when  \texttt{deform = 2}
\end{itemize}

When \texttt{irun = 1}, i.e. the program starts from an existing solution of densities, additional input files \texttt{dens\_n.cwr} and \texttt{dens\_p.cwr} are needed, see
\cref{sec:output_slda}.

\subsubsection{Output} \label{sec:output_slda}
In each self-consistent iterations, the local densities are saved as binary files \texttt{dens\_n\_\#.cwr} and \texttt{dens\_p\_\#.cwr} where \texttt{\#=mod(m,2)} and \texttt{m} is the iteration number. In each file, the datas written and their corresponding offset are (for each proton and neutron)
\begin{table}[htbp]
\center
\begin{tabular}{c|c|c}
\hline\hline

offset (in double) & variable name & meaning \\
\hline
0  & \texttt{nx} & $N_x$\\
1  & \texttt{ny} & $N_y$\\
2 & \texttt{nz} & $N_z$\\
3 & \texttt{dx\_dy\_dz} &  \{$dx, dy, dz$ \}  \\
6 & \texttt{rho} & $n(\vec{r})$\\
6+\texttt{Nxyz} & \texttt{tau} & $\tau(\vec{r})$\\
6+\texttt{2*Nxyz} & \texttt{div\_jj} & $\vec{\nabla} \cdot \vec{J}(\vec{r})$ \\
6+\texttt{3*Nxyz} & \texttt{nu} & $\kappa(\vec{r})$\\
6+\texttt{5*Nxyz} & \texttt{amu} & \{ $\mu,\lambda_x,\lambda_y,\lambda_z,\lambda_{q2}$ \}\\ 
\hline\hline
\end{tabular}

\end{table}

At the end of the program, qpwfs are written into disks as \texttt{wf\_n.cwr} and \texttt{wf\_p.cwr}. A binary file named \texttt{info.slda.solver} is also generated to store the following variables:
\begin{table}[htbp]
\center
\begin{tabular}{c|c|c}
\hline\hline

offset (in double) & variable name & meaning \\
\hline
0-1  & \texttt{nwf\_p, nwf\_n} & number of qpwfs\\   
2-3  & \texttt{amu\_p, amu\_n} &  $\mu_p, \mu_n$\\
4-6  & \texttt{dx,dy,dz} & $dx, dy, dz$ \\
7-9  & \texttt{nx,ny,nz} & $N_x, N_y, N_z$ \\
10   & \texttt{ecut}     & $E_{\mathrm{cut}}$  \\
11   & \texttt{lamda2}    & \{$\lambda_x, \lambda_y, \lambda_z, \lambda_{q2}$\} \\
\hline\hline
\end{tabular}

\end{table}

%

\subsection{TD code}
\subsubsection{Input}
Besides the input file \texttt{info.slda.solver} and wavefunctions \texttt{wf\_*.cwr}, the TD code passes arguments from command line to the main program, which contains the following options
\begin{itemize}
\item
\texttt{-g} number of gpu per node 
\item
\texttt{-t} check-point time (in seconds).
\item
\texttt{-s} total number of steps
\item
\texttt{-i} mode of running: 0 (default) means one-body dynamics; 1 means two-body dynamics (nuclear reaction); 2 means testing run with plane waves as initial wavefunctions.  See more in \cref{sec:test}.
\item
\texttt{-f} type of force (NEDF), the same with static code.
\end{itemize}

\subsubsection{Output}
For every \texttt{loop\_io} time step, the function \texttt{system\_energy} writes into \texttt{results\_td.dat} a line of numbers of the following quantities:

\begin{itemize}
\item
the current time $t$; the total energy $E(t)$; the proton and neutron number $Z(t)$, $N(t)$; 
\item
the center of mass of nucleons $\vec{r}_{\mathrm{cm}}(t)$, 
proton $\vec{r}_{\mathrm{cm},p}(t)$, and neutron $\vec{r}_{\mathrm{cm},n}(t)$; 
\item
the collective flow energy $E_{\mathrm{coll}}(t)$; 
\item
the initial total energy $E(0)$; 
\item
the multiple mass moments $Q_{20}(t), Q_{30}(t), Q_{40}(t)$; 
\item
the average pairing gap $\Delta_p(t)$ and $\Delta_n(t)$;
\item
the energy of density coupling to external field $E_{\mathrm{ext}}(t)$;
\item
the center of mass kinetic energy $E_{\mathrm{cm}}(t)$.
\end{itemize}
Currently \texttt{loop\_io}=100.
%
%

In every 100 time steps, the program opens binary files named \texttt{dens\_all\_n.dat.\#} and \texttt{dens\_all\_p.dat.\#} where $\#$ is the number of time steps. Within the 100 time steps, the densities buffer \texttt{d\_densities} is written into the files with total size \texttt{14*Nxyz*8} bytes for every 10 time steps for each isospin.

In every 500 fm/c, the wavefunctions will be written into disk as \texttt{wf\_n.cwr} and \texttt{wf\_p.cwr} and overwrite the original files.

\section{Usage}
\subsection{Static code}
The compilation of the static code requires the \texttt{FFTW3}, \texttt{ScaLAPACK}, and \texttt{MPI} libraries.  \texttt{OpenMP} options can be turned on, but are not necessary nor optimized. The \texttt{Lustre Utility} library is needed if targeting a Lustre filesystem for I/O. After the modules are loaded or dependencies built, the package of codes can be compiled with the \texttt{Makefile} scripts provided. Users need to modify the compiler and path to various libraries in the makefile as environment variables:
\begin{itemize}
\item
\texttt{COMP} : the compiler, usually \texttt{cc} in default in Cray systems, \texttt{xlc} for IBM systems, \texttt{nvcc} for targeting Nvidia GPUs, etc. Compilers of Open-MPI \texttt{mpicc} and Intel-MPI \texttt{mpiicc} are also applicable.

\item
\texttt{LIBLUT\_INCLUDE\_OPTS}: the path to the \texttt{include} directory of Lustre (if the user links with Lustre i/o functions);

\item
\texttt{LIBLUT\_LINK\_OPTS}: the path to the \texttt{lib} directory of Lustre  (if the user links with Lustre i/o functions);

\item the user will need to also link with the SCALAPACK library available on the system, if it is not by default included.

\end{itemize}

To run the program, one needs to copy the input files in the same directory with the executable.
The job script can be different for different job launchers installed on various clusters/supercomputers, i.e. Cobalt (\texttt{qsub,aprun}), IBM Spectrum Load Sharing Facility (\texttt{bsub,jsrun}), etc. Here we provide an example job script for Slurm scheduler named \texttt{xnslv.slurm} in the package. The running command for the job is simply 
{\small
\begin{verbatim}
srun -n ${NUMBER_OF_CPUS} ./lise-static input.test.txt
\end{verbatim}}

\subsection{TD code}

The compilation of the TD code requires the \texttt{cudatoolkit} (the CUDA Toolkit), \texttt{FFTW3}, and \texttt{MPI} libraries. The \texttt{LUSTRE} library is also needed if the Lustre I/O is enabled. Similar to the static code, the package of TD code can be compiled with the \texttt{makefile} scripts in the package. Besides customizing the environment variables in the static code package, users need to be careful of the option in the compilation of CUDA source code, i.e. \texttt{--gpu-architecture=sm\_60}, which depends on the type of GPUs in the machine. For example, in the template makefile we use \texttt{-gpu-architecture=sm\_60} for NVIDIA Tesla P100 GPUs on Piz-Daint supercomputer. For Titan, which is equipped with NVIDIA Tesla K20 GPUs, one needs to use \texttt{--gpu-architecture=sm\_35}. More details can be found in NVIDIA webpages and relevant user guides.

The template job script of the TD code is \texttt{tdslda.slurm} in the package, which is used on Piz-Daint, with running command
{\small
\begin{verbatim}
srun -n ${NUMBER_OF_GPUS} ./lise-tdslda-gpu ${OPTIONS}
\end{verbatim}}
\noindent
where the environment variables \texttt{OPTIONS} represents the input arguments mentioned in \cref{sec:io}.
On Piz-Daint, each node is equipped with only 1 GPU. For nodes which contain multiple GPUs, users should modify the \texttt{-g} option to specify the number of GPUs per node.

\section{Test cases}\label{sec:test}
We provide two test cases for the code package. All the makefiles, job scripts are written for Summit.  In both cases, the initial qpwfs are generated by the static SLDA code input the provided self-consistent density profiles \texttt{dens\_*.cwr} and \texttt{input.text.txt}. The static calculation involves only one diagonalization of HFB matrix and the qpwfs are written into disk named \texttt{wf\_*.cwr}. The time-dependent calculation is followed by running the TDSLDA code input \texttt{wf\_*.cwr} and \texttt{info.slda\_solver} with appropriate command-line options. 

\subsection{Fission dynamics of \ce{^{240}}{Pu}} \label{sec:pu240}
In this test case we use the SeaLL1 EDF with volume pairing (\texttt{iforce}=7). The size of our simulation box is $30 \times 30 \times 60~\mathrm{fm}^3$ with lattice constant $dx=1.25~\mathrm{fm}$. We use the cubic cutoff in pairing interaction and the number of qpwfs is $n_{\mathrm{wf}}=4N_{xyz} = 110,592$ (neutron and proton). To help users estimate how many GPUs to request to meet the minimum demand of memory, we suggest the following estimation formula
\begin{align}\label{eq:gpuest}
n_\mathrm{{GPU}} > \frac{ n_{\mathrm{wf}} \times (4N_{xyz}) \times 11 \times 16 }{2^{30} \times m_{\mathrm{GPU}}}
\end{align} 
where $m_\mathrm{GPU}$ is the memory (in giga bytes) of each GPU. For Tesla P100 and V100 GPUs, $m_{\mathrm{GPU}} = 16$. In practice users might need to request more $n_\mathrm{{GPU}}$ than the minimum value suggested due to some extra memory cost.
There is no external boosts or potentials added on the qpwfs during the evolution and the compound nucleus will separate into two fragments after a few thousand fm/c. In each time step we track the distance between the center of mass of two fragments, and the evolution will be stopped when the distance reach a value $d_0 \approx 30~\mathrm{fm}$, i.e. the fragments are well-separated. Users can revise the value of $d_0$ by changing the value of variable \texttt{d0} in \texttt{ctdslda.c} source code.

\subsection{Collision of two \ce{^{120}}{Sn}} \label{sec:sn120}
In this test case we use the SLy4 EDF with volume pairing (\texttt{iforce}=1). The size of our simulation box is $25 \times 25 \times 60~\mathrm{fm}^3$ with lattice constant $dx=1.25~\mathrm{fm}$. Two \ce{^{120}}{Sn} nuclei are boosted with $E_\mathrm{cm} = 360~\mathrm{MeV}$ and impact parameter $b=0$ (head-on collision) in $x$-$z$ plane. The $E_{\mathrm{cm}}$ is the sum of the collective kinetic energy of two nuclei at infinite distance. Users can customize $E_{\mathrm{cm}}$ and $b$ by changing the values of variable \texttt{ecm} and \texttt{b} in \texttt{ctdslda.c} source code.

\subsection{Postprocessing and utility codes}
To help users extract useful information from the saved density profiles, we provide utility codes in postprocessing. The \textit{postprocessing} code package, which contains the main function in \texttt{postprocess.c} and other subroutines, reads various local densities from binary files \texttt{dens\_all\_*.dat.\#} and calculates various one-body variables written into text files \texttt{out.dat}, \texttt{outL.dat}, \texttt{outR.dat}, which can be read directly by Python and Matlab for further analysis. The columns of these data files are:
\begin{itemize}
\item
\texttt{out.dat}: 
\begin{itemize}
\item
the current time $t$; 
\item
the total energy $E(t)$; 
\item
the proton and neutron number $Z(t)$, $N(t)$; 
\item
the center of mass of nucleons $\vec{r}_{\mathrm{cm}}(t)$, 
proton $\vec{r}_{\mathrm{cm},p}(t)$, and neutron $\vec{r}_{\mathrm{cm},n}(t)$; 
\item
the collective flow energy $E_{\mathrm{coll}}(t)$;
\item
the quadruple mass moments $Q_{20}(t)$;
\item
the velocity of center of mass $\vect{v}_{\mathrm{cm}}(t)$;
\item
the octuple mass moments $Q_{30}(t)$;
\item
the hexaduple mass moments $Q_{40}(t)$;
\item
the elongation parameter $2Q_{zz}(t)/(Q_{xx}(t) + Q_{yy}(t))$;
\item
triaxial parameters: real and imaginary part of $Q_{21}(t)$ and $Q_{22}(t)$;
\item
the Coulomb energy $E_{\mathrm{Coul}}$ between left and right fragments.
\end{itemize}

\item
\texttt{outL.dat}: the same with \texttt{out.dat}, but for left fragment.
\item
\texttt{outR.dat}: the same with \texttt{out.dat}, but for right fragment.
\end{itemize}

The visualization of the simulation is realized with VisIt \cite{visit}, which is an open source, interactive, scalable, visualization, animation and analysis tool developed by Lawrence Livermore National Laboratory (LLNL).  We provide a utility code \texttt{tdslda2silo.c} to help transfer the original binary files into compatible \textit{silo} files to be read by VisIt. This code needs to be compiled with the \textit{silo} library \cite{silo}.

\section{Performance}
The performance of the static code has been illustrated in Ref.~\cite{jin2017}. Hence,
 we will not discuss the performance of the static code in this paper because it just serves as a utility 
for the TD code and  computational cost of the former one is much smaller than
the latter one for large scale simulations.
Our TD code has been benchmarked on Titan and Summit at OLCF, Oak Ridge, USA and on Piz Daint 
in Lugano, Switzerland. In \cref{tab:titan} we list the time cost of our code on different 
supercomputers. As a measure we have 
used the required computation time per lattice point of one of the components 
of a single qpwf, when performing a complete calculation of all the qpwfs 
\begin{align}
\mathrm{Cost} =\frac{(\# \;\textrm{CU})\times(\textrm{wall-time})}{
(\#\; \textrm{time-steps})\times(\#\; \textrm{PDEs})\times(\#\;\textrm{lattice-points)}} ,
\end{align}
where $\#$ CU stands for the number of computing units, either GPUs in case of the TDSLDA code or CPUs in case of 
other CPU codes in the market for TDHF. The difference of cost is mainly due to the difference of GPU hardwares on 
these computers. The NVIDIA Tesla P100 GPU on Piz-Daint is almost 3x faster than the NVIDIA Tesla K20 
GPU on Titan. The newest NVIDIA V100 GPU on Summit is 30\% faster than P100.

\begin{table}[!h]
\center
\resizebox{1.0\columnwidth}{!}{
\begin{tabular}{ l |  r |  r | r | r | r }
\hline \hline
Code      & CUs  & Computer     & PDEs    &  Lattice               & Cost (sec.)  \\ \hline 
TDSLDA             &   514 & Titan             & 442,368 & $24^2\times48$  & $4.35\times10^{-8}$ \\
TDSLDA             &  240  & Piz Daint      & 442,368 &  $24^2\times48$ & $1.61\times10^{-8}$ \\
TDSLDA-opt  &  240  & Piz Daint      & 442,368 &  $24^2\times48$ & $1.23\times10^{-8}$ \\
TDSLDA             &  240  & Summit        & 442,368 &  $24^2\times48$  & $1.12\times10^{-8}$  \\
TDSLDA-opt   &  240  & Summit        & 442,368 &  $24^2\times48$  & $7.18\times10^{-9}$  \\
\hline \hline
\end{tabular}
}
\caption{\label{tab:titan}Comparison between different existing codes for 
performing TDDFT calculations on a variety of architectures. The TDSLDA code 
demonstrates an almost perfect strong scaling on Piz Daint (Lugano) and Summit 
(Oak Ridge), where further significant optimizations are likely. 
TDSLDA-opt is an optimized version of our GPU code which reduces the number of calls of 
CPU-based routines.}
\end{table}

\begin{figure}[htbp]
\center
\includegraphics[scale=0.5]{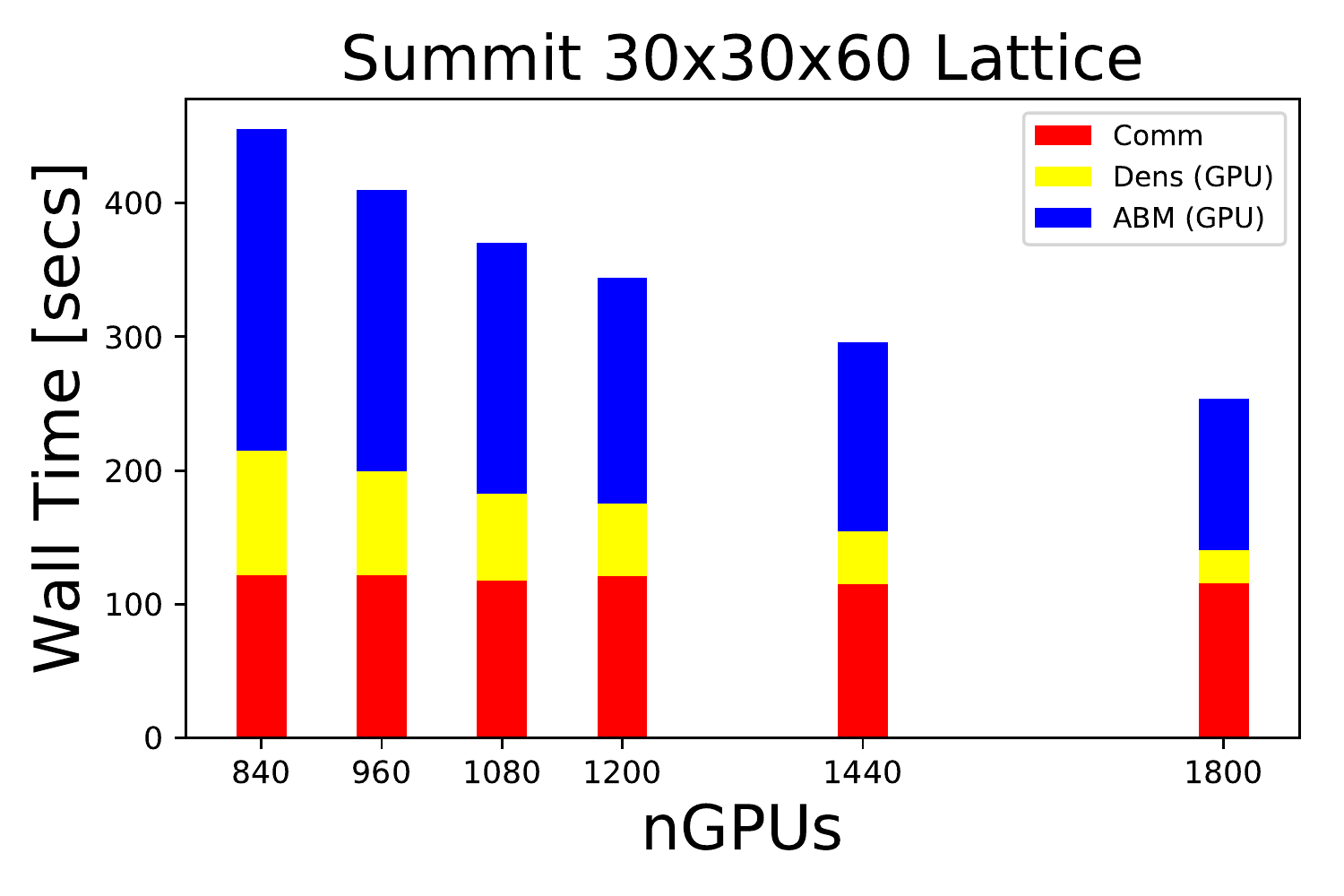}
\includegraphics[scale=0.5]{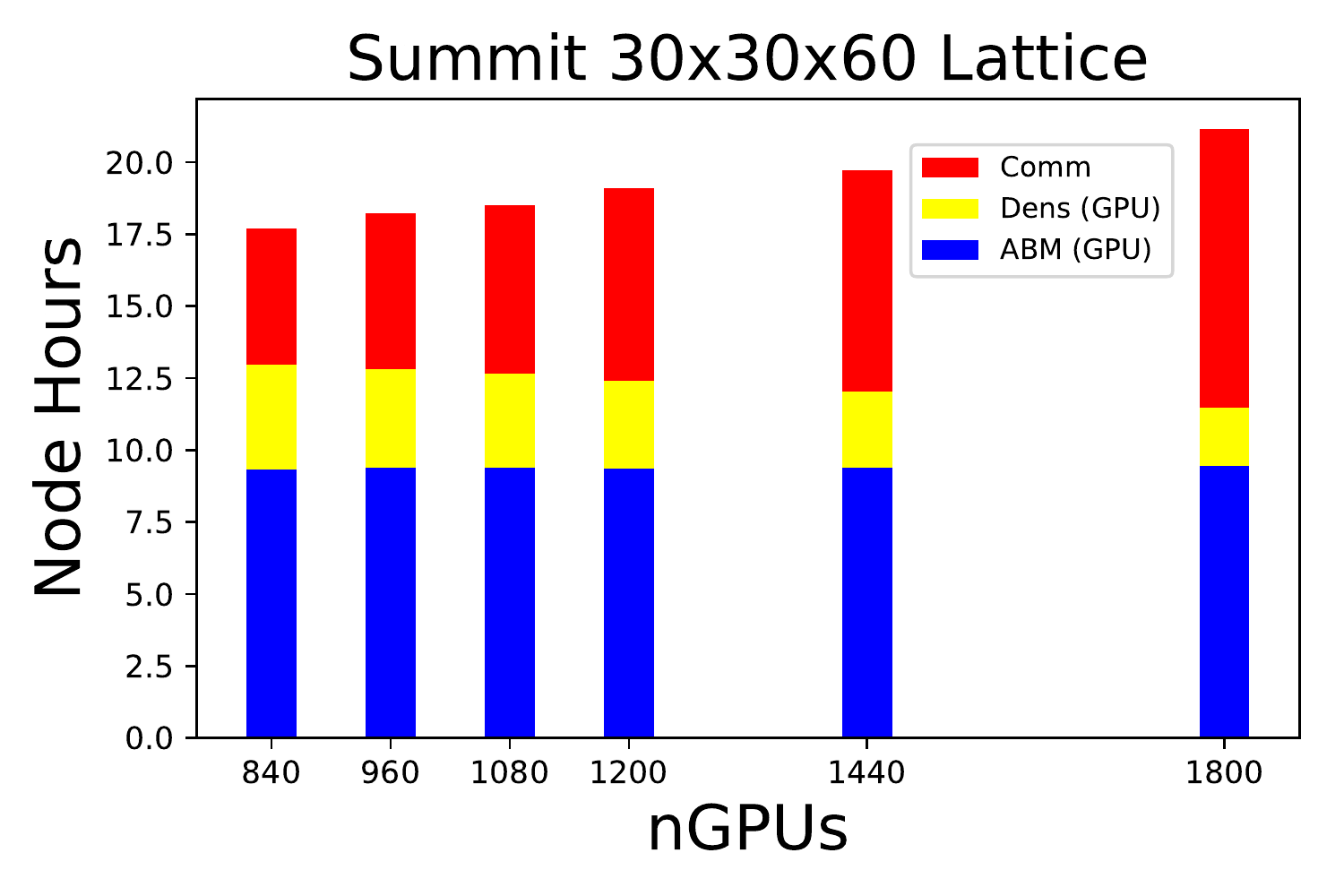}
\caption{\label{fig:node_hours_summit}
The strong scaling capability of the TDSLDA codes on Summit
in the case of fission of $^{236}$U using varying number of nodes.
The cost to perform the pure time evolution is practically constant. Only the 
partial density calculations show a relatively weak dependence on the number of nodes, as
the number of qpwfs to be accounted for per GPU decreases with the number of GPUs 
by approximately a factor of 2 from 840 to 1,800 GPUs. 
The communication between nodes however shows an noticeably increase with 
the number of nodes for more than $\approx 180\dots 200$ nodes. Each node on 
Summit has 6 GPUs.
}
\end{figure}

The scaling properties of different sections in the code are also studied. Here we divide the runtime of 
the TDSLDA code into three parts: 
\begin{itemize}
\item 
ABM: the time cost on the GPU kernels in \texttt{adams\_bashforth\_pm\_gpu()},
\texttt{adams\_bashforth\_cy\_gpu()}, and \texttt{adams\_bashforth\_dfdt\_gpu()}.
\item
Density: the time cost on the GPU kernels in \texttt{compute\_densities\_gpu()}, which includes the calculation
of the gradients and laplaceans of qpwfs and partial local densities.  
\item
Communications: the time cost on the communications between CPU and GPU (\texttt{cudaMemcpy()}),
CPU and CPU (\texttt{MPI\_Allreduce()}).
\end{itemize}
In practice we made a few test runs of TDSLDA code for the same problem described in \cref{tab:titan} up to 400 time steps, 
with different number of nodes on Summit and Piz Daint. In each run, we measure 
the wall-time (in seconds) of each section with the internal 
clock and calculate their corresponding node hours cost on Summit as node hrs. = wall-time (sec.) $\times$ \# nodes / 3600.
As shown in \cref{fig:node_hours_summit}, the cost of running the GPU kernels in ABM and Density parts of the code 
remains almost constant as the number of nodes increases, 
which demonstrates a good strong scaling property. The Comm. part of the calculations 
grows slowly initially and starts to increase only for relatively 
large number of  nodes. 
Clearly the GPU kernels have perfect scaling properties for
 number of GPUs less than 720. Beyond this number, the workload on each GPU 
 is too small and the threads in GPUs are not utilized to full capacity.

We also compared the efficiency of our code with that of the-state-of-the-art 
codes in literature for TDHF calculations~\cite{Maruhn:2014,Umar:2017}, see 
\cref{tab:perf_compare}. The TDHF Sky3D code~\cite{Maruhn:2014} evolves at most 
$\approx 1,000$ PDEs for the collision of two heavy-ions treating pairing 
correlations within the BCS approximation. The wall-time using a number of CPUs 
equal to the number of GPUs in our approach is almost 100x longer for similarly 
sized problems.  We attribute the superior 
performance of the TDSLDA solver to the use of a more efficient while very 
accurate time-integration algorithm, as well as to the use of GPUs. The use of 
highly efficient and precise FFT for the computation of spatial derivatives 
could also be a factor. Since in our calculations we have to manipulate large 
amounts of data, we have taken advantage of fast I/O methods.

\begin{table}[htbp]
\center
\resizebox{1.0\columnwidth}{!}{
\begin{tabular}{ l |  r |  r | r | r | r }
\hline \hline
Code      & CUs  & Computer     & PDEs    &  Lattice               & Cost (sec.)  \\ \hline 
TDSLDA-simp    & 2     & Titan            & 684        &  $20^2\times 60$ &  $7.55\times10^{-8}$ \\
Sky3D~\cite{Maruhn:2014}     &  128  & Titan             &  1,024    & $18^2\times30$  & $3.86\times10^{-6}$ \\
U\&S~\cite{Umar:2017}        &   16  & Linux cluster & 714       & $40^2\times70$  & $8.72\times10^{-5}$ \\ 
\hline \hline
\end{tabular}
}
\caption{\label{tab:perf_compare}Comparison between different existing codes for 
performing TDDFT calculations on a variety of architectures. The TDSLDA code 
demonstrates an almost perfect strong scaling on Piz Daint (Lugano) and Summit 
(Oak Ridge), where further significant optimizations are likely. 
TDSLDA-opt is an optimized version of our GPU code which reduces the number of calls of 
CPU-based routines.
TDSLDA-simp is a simplified and un-optimized version of our GPU code, performing the same type 
of calculations as codes~\cite{Maruhn:2014,Umar:2017} used in literature for 
TDHF+TDBCS simulations.}
\end{table}


\section{Acknowledgment}

We thank many people with whom we had discussions over the years and for their input: 
George F. Bertsch, Yuan-Lung Luo, Piotr Magierski,  Nicolas Schunck, Gabriel Wlaz{\l}owski, Yongle Yu.

The work of AB and SJ was supported by U.S. Department of Energy,
Office of Science, Grant No. DE-FG02-97ER41014 and in part by the NNSA
cooperative agreement DE-NA0003841.  The work of IS was supported by the US Department of Energy through the 
Los Alamos National Laboratory. Los Alamos National Laboratory is operated 
by Triad National Security, LLC, for the National Nuclear Security Administration 
of U.S. Department of Energy (Contract No. 89233218CNA000001). 
IS gratefully acknowledges partial
support of the U.S. Department of Energy through an
Early Career Award of the LANL/LDRD Program and partial support and computational resources provided by the Advanced Simulation and Computing (ASC) Program.
The work of KR was partially supported by the Exascale Computing Project (grant no. 17-SC-20-SC), a collaborative effort of two US DOE organizations (Office of Science and the National Nuclear Security Administration), and by the US Department of Energy through the 
Pacific Northwest National Laboratory. Pacific Northwest National Laboratory is operated by Battelle Memorial Institute for the U.S. Department of Energy under Contract DE-AC05-76RL01830. 
The TDSLDA calculations have been
performed at the OLCF Summit, Titan, and Jaguar,  and CSCS Piz Daint, and for generating
initial configurations for direct input into the TDSLDA code at OLCF
Titan and Summit and Edison at NERSC. This research used resources of the Oak Ridge
Leadership Computing Facility, which is a U.S. DOE Office of Science
User Facility supported under Contract No. DE- AC05-00OR22725 and of
the National Energy Research Scientific computing Center, which is
supported by the Office of Science of the U.S. Department of Energy
under Contract No. DE-AC02-05CH11231.  We acknowledge PRACE for
awarding us access to resource Piz Daint based at the Swiss National
Supercomputing Centre (CSCS).
This work was also supported by "High Performance Computing 
Infrastructure" in Japan and a series of simulations 
were carried out on the Tsubame 3.0 supercomputer at Tokyo Institute of Technology.
This research also used resources provided by the Los Alamos National Laboratory 
Institutional Computing Program.






\bibliographystyle{elsarticle-num}
\bibliography{CPC_slda}







\end{document}